\documentstyle[psfig,twocolumn,aps,amstex,floats,graphicx]{revtex}

\tighten
\begin{document}

\renewcommand{\topfraction}{1.0}
\renewcommand{\textfraction}{0.0}

\draft

\title{\bf High Order Perturbation Theory for Spectral Densities of
  Multi-Particle Excitations: $\mathbf{S=\frac{1}{2}}$ Two-Leg Heisenberg
  Ladder}  

\author{Christian Knetter, Kai P. Schmidt and G\"otz S. Uhrig}

\address{Institut f\"ur Theoretische Physik, Universit\"at zu
  K\"oln, Z\"ulpicher Str. 77, D-50937 K\"oln, Germany\\[1mm]
  {\rm(\today)} }

\maketitle

\begin{abstract}
We present a high order perturbation approach to quantitatively calculate
spectral densities in three distinct steps starting
from the model Hamiltonian and the observables of interest. The
approach is based on the perturbative continuous unitary
transformation introduced previously. It is conceived to work
particularly well in models allowing a clear identification of the
elementary excitations above the ground state. These are then viewed
as quasi-particles above the vacuum. The article
focuses on the technical aspects and includes a discussion of 
series extrapolation schemes. The strength of the method is
demonstrated for $S=1/2$ two-leg Heisenberg ladders, for which results are
presented. 
\end{abstract}


\narrowtext

\section{A. Introduction}
Spectroscopic measurements provide important insights in the
microscopic structure of solids. Generic spectroscopic data contains
information about energy bands, associated density of states, matrix
elements and selection rules. 
From the theoretical point of view this information is indispensable
in the process of formulating and testing appropriate microscopic
models. However, a quantitative comparison of the theoretical and the
experimental data strongly depends on the power of the method
used to calculate the properties of the microscopic model. In this context high
order perturbation theory has proven to be a versatile and flexible
tool. Especially in the field of spin models a variety of perturbation
techniques is in use (for an overview see Ref.~\cite{gelfa00}). Most of
them concentrate on the calculation of one-particle energies and in some
circumstances on the associated spectral
weights~\cite{singh95,singh99,weiho01}. However, so far the important
information contained in the line 
shapes of spectroscopic data, i.e., the model's
spectral densities associated with the experimental observables, has
not been exploited. The need for a {\it quantitative} calculational
tool closing this gap is apparent. 

In the last couple of years there has been considerable progress in
the field of high order perturbation theory. For a long time the
methods were restricted to ground state energies and one-particle
energies. Just recently we were able to quantitatively calculate {\it
  two}-particle excitation energies (bound states) in various spin
models~\cite{uhrig98c,knett00b}. These calculations were based on the
perturbative continuous unitary transformation (CUT) method introduced
previously~\cite{knett99a}. This technique was used successfully
for low-energy calculations in various spin models
before~\cite{knett99a,knett00a,knett00c}. The key point is to
construct the transformation such that the resulting effective
Hamiltonian is block-diagonal with respect to the number of particles.
The linked cluster series expansion, an established high order
perturbation method, has been shown
lately to be also well suited for calculating two-particle
energies~\cite{trebs00,weiho00a}. The basic idea is again
the application of an orthogonal transformation which 
is designed to achieve a block-diagonal effective Hamiltonian.

In Ref.~\cite{knett03a} we presented an extension of the CUT method
allowing a systematic high order perturbation theory for
observables. In the present article we use the results of
Ref.~\cite{knett03a} to calculate spectral densities of
multi-particle excitations quantitatively. The method allows to obtain
the complete spectral information for experimental relevant
observables without any 
finite size restrictions. The results are exact in the sense of the 
thermodynamic limit, i.e., each order can be calculated for the
thermodynamic limit. By truncating the series expansion at a (high) maximum
order we restrict to dynamic processes for which the involved
particles interact within a certain finite distance to each
other. Obtaining higher orders amounts to allowing larger
distances. Hence, the scheme can be expected to work particularly well
in systems with short correlation lengths.

Results for various spin systems have been reported in earlier
publications 
(Refs.~\cite{knett01a,schmi01a,knett02b,windt01,gruni02c,schmi03a,schmi03b,schmi03c,knett03b}). The article on hand 
gives the technical details necessary to apply the method.
We like to mention that the linked cluster method mentioned above was
recently extended to allow for calculating spectral
densities, too~\cite{zheng03a}. It thus constitutes an alternative
approach.

To illustrate our approach we consider the $S=1/2$
antiferromagnetic two-leg Heisenberg ladder as an interesting and
comprehensive testing ground. The Hamiltonian reads 
\begin{align}
  \label{H_start}
 H&(J_{\perp},J_{\parallel})=
 J_{\perp}H_{\perp}+J_{\parallel}H_{\parallel}\\\nonumber      
 =&\sum_{i}\left [ J_{\perp}{\mathbf S}_{1,i} {\mathbf S}_{2,i}  
 \!+\!\!\! \ J_{\parallel}\left({\mathbf S}_{1,i} {\mathbf
  S}_{1,i+1}+{\mathbf S}_{2,i} {\mathbf S}_{2,i+1}\right)\right]\ , 
\end{align}
where $i$ denotes the rung and $1,2$ the leg.

In the broad field of spin liquid systems there has been an
ongoing theoretical interest in the spin ladder and its extended
versions~\cite{barne93,barne94,gopal94,totsu95,brehm96,weiho96a,sylju97,sush98,natsu98,weiho98a,pieka98,brehm99,kotov99,breni00,singh00,weiho00a,zvyag01}.
The model is realized in a number of substances~\cite{johns00} and there
is a large amount of experimental data available, see
e.g. Refs.~\cite{kojim95,schwe96,eccle96,kuma97,hamma97,sugai99,matsu00b,konst01}. 
Additionally, the experimental evidence for 
superconductivity in Sr$_{0.4}$Ca$_{13.6}$Cu$_{24}$O$_{41.84}$ under
pressure~\cite{uehar96} has intensified the interest.

\setcounter{section}{0}

\section{B. Article Outline}
The basic concept of our approach to spectral densities is
as follows. For a given observable $\mathcal{O}$ the $T=0$ spectral
density is calculated from 
\begin{equation}
  \label{spec_1}
  {\mathcal S}(\omega)=-\frac{1}{\pi}{\rm Im} {\mathcal G}(\omega) \ ,
\end{equation}
where ${\mathcal G}(\omega)$ is the retarded zero temperature
Green function 
\begin{equation}
  \label{green_1}
  {\mathcal G}(\omega)=\langle\psi_0|{\mathcal O}^{\dagger}(\omega
  -(H-E_0)+i0+)^{-1} {\mathcal O}|\psi_0 \rangle\ .
\end{equation}
Here $\psi_0$ is the ground state and $E_0$ is the ground state energy
of the system. Since the expectation values of quantum mechanical
observables do not change under unitary transformations,
the Green function ${\mathcal G}$, and thus ${\mathcal S}$, will not
be altered if the operators $H$ and ${\mathcal O}$ and the state
$|\psi_0\rangle$ appearing in ${\mathcal G}$ are substituted by the
{\it effective} operators (state) $H_{\rm eff}$ and ${\mathcal O}_{\rm
  eff}$ ($|\psi_{0,\rm eff}\rangle$) obtained from the CUT method. Our
procedure can be divided into three steps:  
\begin{enumerate}
\item  
Use the CUT to derive an effective Hamiltonian $H_{\rm eff}$
unitarily linked to $H$.\\
\item 
Use the {\it same} transformation to derive the effective
  observable ${\mathcal O}_{\rm eff}$ from some initial observable
  ${\mathcal O}$ of interest.\\
\item 
Evaluate Eq.~(\ref{green_1}) for the effective operators in
  terms of a continued fraction.
\end{enumerate}
Now, the key ingredient of our approach is to identify suitable
(quasi-)particles which can be used to describe the $T=0$
physics of the system under study. We then proceed and construct the
perturbative unitary transformation such that the resulting effective
Hamiltonian $H_{\rm eff}$ {\it conserves} the number of these
particles. Let $Q$ be the operator which counts the number of
particles. Then the conservation of the number of quasi-particles reads
\begin{equation}
  \label{H_Q_comm_I}
  [H_{\rm eff},Q]=0\ .
\end{equation}
In Sect.~II we describe in detail how this idea can be put to use by
illustrating the procedure for the spin ladder example.

Applying the same transformation to observables ${\mathcal O}$
different from the Hamiltonian
leads to effective observables ${\mathcal O}_{\rm eff}$ not conserving
the number of particles in 
general. Their action on the ground state, as needed in the
evaluation of the Green function~(\ref{green_1}), is
characterized by the number of particles they inject, i.e., by the number
of elementary excitations they excite. We thus
decompose ${\mathcal O}_{\rm eff}$ into operators injecting none, one,
two and so on particles in the system. In Sect.~III we again use the
ladder example to illustrate the practical realization of this concept
for experimentally relevant observables. 

Once the effective Hamiltonian and the observables are obtained they are
inserted into the Green function. Sect.~IV features a detailed
description of how the resulting expression is manipulated to extract
the corresponding (energy- and momentum-resolved) spectral
densities. Again, the one-, two- and 
more-particle contributions to the total spectral density can be treated
separately leading to a simple and comprehensive physical picture in
the end.

In Sect.~V we address the problem of series extrapolation, an
inevitable difficulty in perturbative approaches. Following our
approach, a very large number of quantities has to 
be extrapolated simultaneously. This poses a difficulty which cannot
always be tackled by standard techniques. We introduce a robust
extrapolation scheme based on optimized perturbation
theory~\cite{steve81} and show how it can
be applied to extend the range of the perturbative results.

The article is summarized in Sect.~VI.

\section{Transformation of the Hamiltonian}
\label{H_transformation}
We start by briefly explaining how the effective Hamiltonian $H_{\rm
  eff}$ is constructed from the initial ladder
  Hamiltonian~(\ref{H_start}). (Other examples for this procedure can
  be found in Refs.~\cite{knett99a,knett00c} for instance.)
In the subsequent sections II.A through II.C we illustrate in
  detail how the zero-, one- and two-particle energies are calculated
  from $H_{\rm eff}$. 

Let us assume that the initially given Hamiltonian $H$ can be formulated as
perturbative problem 
\begin{equation}
  \label{U_and_V}
  H=U+xV\ .
\end{equation}
In its present formulation (extensions are possible) the perturbative
CUT method relies on two prerequisites calling for a band-diagonal problem
as starting point:
\begin{enumerate}
\item[(A)] The unperturbed Hamiltonian U must have an
  equidistant spectrum bounded from below. The difference between two
  successive levels is called an energy quantum or (quasi-)particle
  and we identify $Q=U$.
\item[(B)] There is a number ${\mathbb N} \ni N>0$ such that the perturbing
  Hamiltonian $V$ can be written as $V=\sum_{n=-N}^{N}T_n$ where $T_n$
  increments (or decrements, if $n<0$) the number of energy quanta by
  $n$: $[Q,T_n]=nT_n$.
\end{enumerate}
We now show that the initial ladder Hamiltonian~(\ref{H_start}) meets
these requirements. We reformulate the ladder problem according to
\begin{equation}
\label{H_pert}
  \frac{H(x)}{J_{\perp}}=H_{\perp}+xH_{\parallel},
\end{equation}
with $x=J_{\parallel}/J_{\perp}$ as perturbation parameter,
$H_{\perp}=H(1,0)$ and $H_{\parallel}=H(0,1)$. We assume $J_{\perp}$
to be antiferromagnetic and set $J_{\perp}=1$ henceforth. The
limit of isolated rungs is the limit for which our perturbative
treatment is controlled.

The ground state of the unperturbed part $H_{\perp}$ is the product
state with singlets on all rungs. A first excited state is 
a single rung excited to a triplet. There are $3L/2$ such elementary
triplet excitations if $L$ is the number of spins. The energetically
next higher 
state is given by two rung-triplets and so on.  The operator $H_{\perp}$
simply counts the number of rung-triplets and it is easily
verified that condition (A) is fulfilled. 

For the rest of this article we identify $Q=H_{\perp}$, i.e., the
elementary excitations of the unperturbed part (rung-triplets) serve
as (quasi-)particles in our treatment of the ladder system. Since we
prefer the particle picture we call these elementary
excitations {\it triplons} \cite{schmi03a}. 
They are total spin $S=1$ excitations appearing in 
three different variants ($S^z$-components -1,0,1) in contrast to magnons
which have two variants only ($S^z \pm 1$) in a phase
of broken spin symmetry. Both, triplons and magnons, must
be distinguished from spinons, which are $S=1/2$ excitations. 
Whenever we refer to the ladder system we will use the term
triplon. In more general discussions we retreat to the term
(quasi-)particle.

As soon as we turn on the inter-rung interaction ($x>0$) the
triplons become dressed particles. The central idea of the CUT approach is
to map the initial problem onto an effective Hamiltonian for which the
simple triplon-states, originally defined for the unperturbed part,
can be used to calculate all energy levels of the system.

We proceed and analyze the action of the perturbing part $H_{\parallel}$
on the triplon-states. Let $|n\rangle$ denote a state with $n$ rungs
excited to triplets ($n$-triplon state),
i.e., $H_{\perp}|n\rangle=n|n\rangle$. Then 
\begin{eqnarray}
\label{H_decomp}
H_{\parallel}&=&T_{-2}+T_0+T_2\ ,\mbox{  with}\\\nonumber
T_i|n\rangle &\sim& |n+i\rangle\ \mbox{  and}\\\label{T_local}
T_{0,\pm 2}&=&\sum_{\nu} {\mathcal T}_{0,\pm 2}(\nu)\ ,
\end{eqnarray}
where $\nu$ denotes pairs of adjacent rungs. The index $\nu$ can also
be viewed to count the bonds connecting adjacent rungs. 
The action of the local operators ${\mathcal T}_{0,\pm 2}(\nu)$ on
neighbouring rungs is given in Table I. Condition (B) is fulfilled
with $N=2$. There appear no
$T_{\pm 1}$ in $H_{\parallel}$. 
\begin{table}[h]
\begin{tabular}{|ccc|}
&$2{\mathcal{T}}_{0}$&\\
\hline\hline
$|t^{0,\pm1},s\rangle$ & $\longrightarrow$ & $|s,t^{0,\pm1}\rangle$\\
$|t^0,t^{\pm1}\rangle$ & $\longrightarrow$ & $|t^{\pm1},t^0\rangle$\\
$|t^{\pm1},t^{\pm1}\rangle$ & $\longrightarrow$ & $|t^{\pm1},t^{\pm1}\rangle$\\
$|t^{\pm1},t^{\mp1}\rangle$ & $\longrightarrow$ & $|t^0,t^0\rangle-|t^{\pm1},t^{\mp1}\rangle$\\
$|t^0,t^0\rangle$ & $\longrightarrow$ & $|t^1,t^{-1}\rangle+|t^{-1},t^1\rangle$\\
\hline
\hline
&$2{\mathcal{T}}_{ 2}$&\\
\hline\hline
$|s,s\rangle$ & $\longrightarrow$ & $|t^0,t^0\rangle-|t^{1},t^{-1}\rangle-|t^{-1},t^1\rangle$
\end{tabular}
\caption{\label{matrix_tabI} Action of the operators 
${\mathcal{T}}_{i}$ as defined by Eq.~(\ref{T_local}) on
product states of adjacent rungs. Singlets are denoted by $s$ and
triplons by $t^i$ where the superscript indicates the magnetic quantum
number. The remaining
matrix elements can be found by using ${\mathcal
 T}_{n}^{\dagger}={\mathcal T}_{-n}^{}$.}
\end{table}

The perturbative CUT is engendered by introducing an auxiliary
variable $\ell \in [ 0,\infty]$. The CUT gives rise to the flow
equation (details in Ref.~\cite{knett99a}) 
\begin{equation}
  \label{flow_eq}
  \frac{\partial H(x;\ell)}{\partial \ell}=[\eta(x;\ell),H(x;\ell)]\ ,
\end{equation}
which controls the flow of the Hamiltonian in the transformation process.
We fix $H(x;0)=H(x)$ and define $H_{\rm eff}(x):=H(x,\infty)$.

 As shown in Ref.~\cite{knett99a} the best
  choice for the infinitesimal unitary generator is (${\rm sgn}(0)=0$)
  \begin{equation}
    \label{gen}
    \eta_{i,j}(x;\ell)={\rm sgn} (Q_i-Q_j)H_{i,j}(x;\ell)\ ,
  \end{equation}
where the matrix elements $\eta_{i,j}$ and $H_{i,j}$ are given in the
eigen basis $\{|n\rangle\}$ of $Q=H_{\perp}$. In the
limit $\ell \rightarrow \infty$  
generator~(\ref{gen}) eliminates all parts of $H(x;\ell)$ changing the number
of particles, i.e., $[H_{\rm eff},H_{\perp}]=0$, and keeps the
flowing Hamiltonian (intermediate $\ell$)
band-diagonal~\cite{knett99a}. The vanishing 
commutator expresses the fact that $H_{\rm eff}$ is block-diagonal
with respect to the number of particles. 

A perturbative realization of the transformation yields
the effective Hamiltonian as operator series expansion
\begin{equation}
  \label{H_eff}
  H_{\rm eff}(x) = H_{\perp} +\sum_{k=1}^{\infty}x^{k} 
\sum_{|\underline{m}|=k, M(\underline{m})=0} C(\underline{m}) 
T(\underline{m})\ .
\end{equation}
Here $\underline{m}$ is a
vector of dimension $|\underline{m}|=k$ of which the 
components are elements of  $\{\pm N, \pm (N-1),\ldots \pm 1, 0\}$. In
the ladder case we have $N=2$ and $T_1=T_{-1}\equiv 0$
(cf.~Eq.(\ref{H_decomp})). The 
operator products $T(\underline{m})$ are defined by 
$T(\underline{m})=T_{m_1}T_{m_2}\cdots T_{m_k}$, with $T_{m_i}$ as
given in Eq.~(\ref{T_local}); $k$ is the
order of the process and $M(\underline{m}):=\sum m_i=0$ signifies 
that the sum of the indices vanishes which reflects the
conservation of the number of particles. Thus the action of
$H_{\rm eff}$ can be viewed as a weighted sum of virtual excitation
processes $T(\underline{m})$ in each of which the particle number is
conserved. The coefficients $C(\underline{m})$ can be calculated as
fractions of integers (in the ladder case up to order $k=15$). The
effective Hamiltonian is thus an exact series expansion up to some
maximum order. 

We want to emphasize that the effective Hamiltonian $H_{\rm
  eff}$ with known coefficients $C(\underline{m})$ 
can be used straightforwardly in all perturbative problems that meet
conditions (A) and (B). The coefficients $C(\underline{m})$ will be
made available electronically on our
web pages~\cite{homepages}.

The action of the effective Hamiltonian~(\ref{H_eff}) on the states of
interest is calculated on a computer. In the
following subsections we illustrate how we obtain perturbative results
for the ground state energy, the one-triplon and the two-triplon
energies for the spin ladder from $H_{\rm eff}$.

On general grounds we showed previously~\cite{knett03a} that $H_{\rm
  eff}$ decomposes into a sum of {\it irreducible} $n$-particle
operators $H_n$ 
\begin{equation}
  \label{H_eff_Zerlegung}
  H_{\rm eff}=\sum_{n=0}^{\infty}H_n\ .
\end{equation}
For the problem on hand the operator $H_n$ measures $n$-triplon
energies no matter how many triplons are present as long as there are
at least $n$ triplons. On states containing less than $n$ triplons the
action of $H_n$ is zero. The matrix elements of $H_n$ are extensive
quantities. By exploiting the linked cluster theorem they can
therefore be calculated perturbatively for the 
infinite system on finite minimum clusters, which are just large enough to
perform the calculations without finite size effects. More details can
be found in Ref.~\cite{knett03a}.

The $H_n$ are calculated recursively from $H_{\rm eff}$ starting with
$H_0$ (Eqs.~(9) in Ref.~\cite{knett03a}). In this way,
Eq.~(\ref{H_eff_Zerlegung}) stands for a systematic
energy-calculation scheme. One starts by
calculating the ground state energy ($H_0$) and proceeds by
calculating one-triplon energies ($H_1$). True two-triplon
interactions can be calculated by including $H_2$ and so on. A
detailed description of this issue, in particular 
how the $H_n$ are defined, is given in Ref.~\cite{knett03a}. 

In the following three subsections
we address the calculation of $H_0$, $H_1$ and $H_2$ for the spin
ladder system separately. All necessary computational details are
presented. Particular attention is paid to the choice of minumum
clusters for the ladder system. The aim is to offer a worked example
for the interested reader.

\subsection{Zero Triplon: $H_0$}

Let $|0\rangle$ denote the triplon vacuum. This is the state where all
rungs are occupied by singlets. Clearly, $|0\rangle$ is the ground
state of $H(x=0)=H_{\perp}$. The one-triplon gap separates the
corresponding ground state energy from the first excited level. In
Ref.~\cite{heidb02b} we showed on general grounds that the particle vacuum 
$|0\rangle$ remains the ground state of $H_{\rm eff}$ for finite
$x$ unless a 
phase transition occurs (e.g. a mode softening at some critical value
$x_c$). For the ladder system in particular, one observes that the
one-triplon gap decreases on increasing $x$ but stays finite for all
$0<x<\infty$~\cite{dagot96,shelt96,greve96}. There are no phase
transitions in this range and 
$|0\rangle$ remains the ground state. Since $H_{\rm eff}$
conserves the number of triplons we conclude that
$\langle 0|H_{\rm eff}(0<x<\infty)|0\rangle$ is the ground state energy. The
point $x=\infty$ 
is a singular point at which the two legs of the ladder
decompose into two decoupled gapless spin 1/2 chains.

Since the action of $H_0$ on $|0\rangle$ coincides with the action of
$H_{\rm eff}$ on this state (see Ref.~\cite{knett03a}), every order of
the ground state energy per site $\epsilon_0$ can be calculated in the
thermodynamic limit on a finite minimum cluster by 
\begin{equation}
  \epsilon_0=\langle 0| H_{\rm eff}|0\rangle/(2N)\ ,
\end{equation}
where $N$ is the number of rungs used in the minimum cluster.

\begin{figure}[h]
  \begin{center}
    \includegraphics[width=5cm]{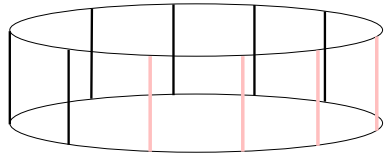}
    \caption{\label{PBC} A periodically closed cluster of ten
      rungs. $H_{\rm eff}$ to third order connects at maximum four
      neighbouring rungs by activated bonds (see text). The connected
      rungs are printed in grey.}
  \end{center}
\end{figure}

We now specify the minimum cluster. At first, it is clear that we need
a {\it closed} ladder segment. This
ensures that there are no end rungs, which are linked to the
cluster by one inter-rung bond only. They would not contribute the same
amount of energy as the fully linked rungs in the middle of the
cluster. Fig.~\ref{PBC} shows a
cluster of the ladder system which has been closed to a ring. 

A close inspection of
Eqs.~(\ref{H_eff}) and (\ref{T_local}) and Tab.~\ref{matrix_tabI}
shows that $H_{\rm eff}$ connects a maximum of $l+1$ rungs on a finite
cluster of $N$ rungs in $l^{\rm th}$ order. In other words: A maximum 
of $l$ bonds between neighbouring rungs can be {\it activated} in
$l^{\rm th}$ order. A bond $\nu$ is said to be activated, if a part of
$H_{\rm eff}$, i.e., the specific local operator ${\mathcal T}_n(\nu)$
in $T_n=\sum_{\nu}{\mathcal T}_n(\nu)$ of $H_{\rm eff}$ (see
Eq.~(\ref{T_local})), has acted on the two rungs connected by $\nu$.

The linked cluster theorem states that only those processes induced
by the $T(\underline{m})$ of $H_{\rm eff}$ contribute to the ground
state energy (and all other extensive quantities), in which all
activated bonds are linked. Processes involving disconnected
active-bond distributions 
cannot contribute. 
\begin{figure}[h]
  \begin{center}
    \includegraphics[width=8.6cm]{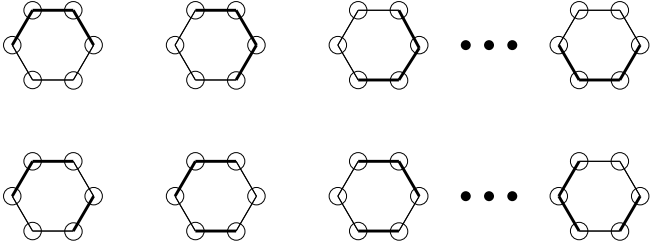}
    \caption{\label{lct_erklaerung} A closed ladder-segment of six
      rungs. Rungs are
      depicted by circles and (active) bonds between rungs by (thick)
      solid lines. In a process of order $l=3$ a 
      maximum of 3 bonds can be active. On a closed cluster of $N=6$
      there are 6 possibilities to arrange {\it linked} bonds (top
      row). One clearly sees that this number grows linearly in
      $N$. The given example of 3 {\it disconnected} active bonds
      (bottom row) has 12 possibilities, which would lead to a
      super-extensive contribution $\propto N^2$ 
      to the extensive quantity under
      study. Thus they do not contribute.} 
\end{center}
\end{figure}
The basic argument is sketched in
Fig.~\ref{lct_erklaerung}. This means in our case, that a cluster of
$l+1$ rungs is sufficient to calculate the $l^{\rm th}$ order contribution 
avoiding {\it wrap-arounds} as indicated in Fig.~\ref{Connect}.
\begin{align}
  \label{min_clust}\nonumber
  &\mbox{Minimum number of rungs to calculate}\\\nonumber
  &\mbox{the $l^{\rm th}$ order contribution to $\epsilon_0$ in
    the}\\\nonumber 
  &\mbox{thermodynamic limit}\\
  &=l+1\ . 
\end{align}
Once the minimum cluster is specified it is straight forward
to calculate $\epsilon_0$. The action of $H_{\rm eff}$ on
$|0\rangle$, which we 
have calculated up to 14$^{\rm th}$ order, on a closed cluster
containing 15 rungs is implemented on a computer. Details of this
procedure can be found in Ref.~\cite{knett99a}. Since $H_{\rm eff}$
conserves the number of triplons we have $H_{\rm eff}|0\rangle\sim
|0\rangle$. The constant of proportionality is the ground state energy
$E_0$ of the 15-rung cluster. The ground state energy per spin is
finally given by $\epsilon_0=E_0/30$. The result is a
14$^{\rm th}$ order polynomial in $x$. It is the exact energy of the
infinite system to the given order 
\begin{align}
  \nonumber
 \epsilon_0&=-\frac{3}{8} -\frac{3}{16}{x}^{2}-{\frac {3}{32}}\,{x}^{3}+{\frac
   {3}{256}}\,{x}^{4}+{\frac {45}{512}}\,{x}^{5}+{\frac
   {159}{2048}}\,{x}^{6}\\\nonumber
 &-{\frac {879}{32768}}\,{x}^{7}-{\frac
   {4527}{32768}}\,{x}^{8}-{\frac {248391}{2097152}}\,{x}^{9}+{\frac
   {336527}{4194304}}\,{x}^{10}\\\nonumber
 &+\!{\frac
   {117840599}{402653184}}\,{x}^{11}\!+{\frac
   {175130171}{805306368}}\,{x}^{12}\!-{\frac
   {58290422737}{231928233984}}\,{x}^{13}\\\label{epsi_0_plain}
 &-{\frac
   {246296576249}{347892350976}}\,{x}^{14}\ . 
\end{align}
The coefficients are fractions
of integers and therefore free from rounding errors. 
Our findings agree with
the numerical results given by Zheng et al.~\cite{weiho98a}. 
\begin{figure}[h]
  \begin{center}
    \includegraphics[width=8.6cm]{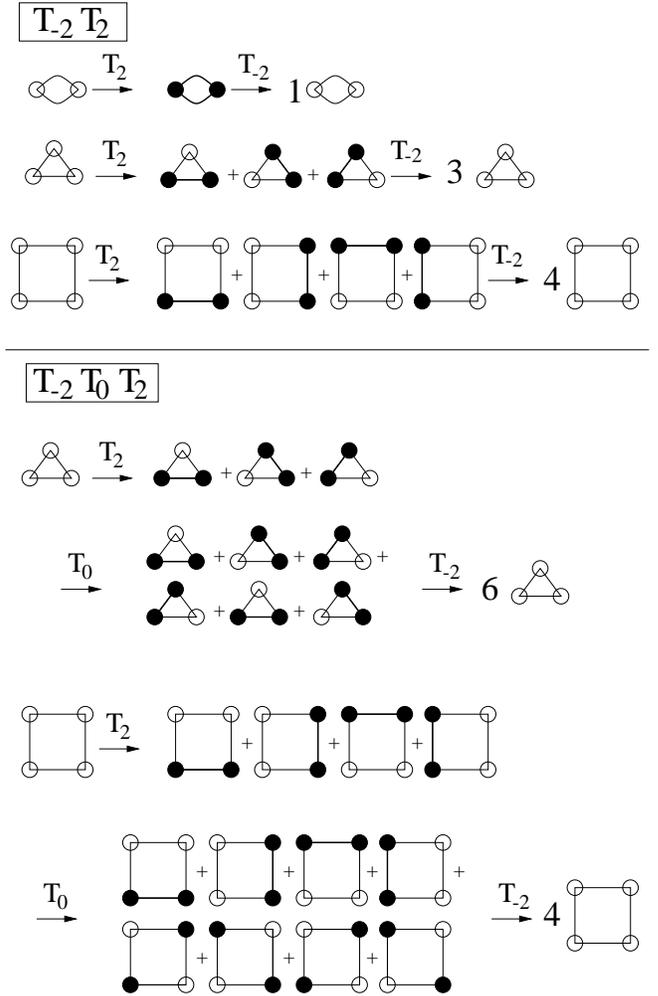}
    \caption{\label{Connect} Symbols as in Fig.~\ref{lct_erklaerung};
      here: rung-singlets (-triplons) are denoted by open (filled)
      circles. In second order only $T_{-2}T_2$ and in third order
      only $T_{-2}T_0T_2$ contribute to $\epsilon_0$. The action of
      these operators on the ground state (always to the left) is
      shown step by step. The upper part shows that in second order
      the ground state energy per site becomes independent of the
      cluster-size, if the cluster contains more than two rungs. The
      lower part shows that wrap-arounds are possible, if the cluster
      is too small: For the third order contribution to $\epsilon_0$ a
      cluster containing three rungs is undersized. The $T_0$ operator
      in the middle of the process can break up one triplon-pair by
      moving one of the triplons away by one rung. On the three-rung
      cluster it joins back the remaining triplon from the other side
      (wrap-around), which would not be possible in the thermodynamic
      limit. This process can be avoided by adding one extra rung as
      buffer, as depicted in the bottom process. Note that $T_{-2}$
      destroys two triplons only if they are neighbours. Thus states
      with diagonally arranged triplons do not contribute. Remembering
      that one 
      always ends with the ground state $|0\rangle$ if one starts
      from $|0\rangle$ ($H_{\rm eff}$ conserves the number of
      triplons!), it is clear that this argument works for all
      possible processes $T(\underline{m})$ in $H_{\rm eff}$.} 
\end{center}
\end{figure}

\begin{figure}[h]
  \begin{center}
    \includegraphics[width=8.6cm]{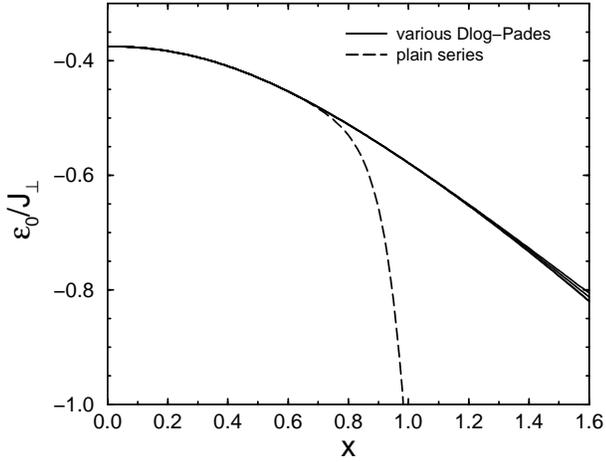}
    \caption{\label{E0_vergl} Ground state energy per spin as function
      of $x$. The plain series result~(\ref{epsi_0_plain}) is depicted
      as dashed line. Four different  Dlog-Pad\'e approximants ([7,6],
      [8,5], [5,8] and [6,7]) are
      shown as solid lines.}
\end{center}
\end{figure}

The polynomial~(\ref{epsi_0_plain}) is depicted as dashed line in
Fig.~\ref{E0_vergl}. The solid 
lines correspond to four different Dlog-Pad\'e
approximants~\cite{domb83_pade} of this quantity and constitute a
reliable extrapolation. The plain series result can be trusted up to
$x\approx 0.7$. 
Since
each rung can be in four different states ($s,t^1,t^0,t^{-1}$) the
Hilbert space has dimension $4^{15}=2^{30}$. Thus the computer
calculations used about 1 GByte. They took about 20h. The polynomial will
be made available on our home pages~\cite{homepages}.

\subsection{One-Triplon Dispersion: $H_1$}
We define $|i\rangle$ to denote the eigen state of
$H_{\perp}$ with one triplon on rung $i$ and singlets on all other
rungs. The magnetic quantum number $m$ of the triplon at rung $i$ is
of no importance in the following considerations, since 
$H_{\rm eff}$ conserves $m$ and the total spin $S$
(cf. Tab.~\ref{matrix_tabI}). Thus it is not denoted explicitly.

Since $H_{\rm eff}(x)$ conserves the number of triplons the action of
$H_{\rm eff}(x)$ on the state $|i\rangle$ is a hopping of the
triplon. We define the hopping coefficients
\begin{equation}
\label{hopp_amp}
a^{\rm cl}_{i;j}(x)=\langle i|H_{\rm eff}(x)|j\rangle\ .
\end{equation}
The superscript ${\rm cl}$ indicates that the hopping coefficient might
depend on the cluster on which it was calculated.

The hopping coefficients $t_{i;j}$ of the irreducible one-particle
operator $H_1$ read (see Eqs.9 in Ref.~\cite{knett03a}) 
\begin{equation}
  \label{hopp_amp_mod}
  t_{i;j}=\langle i|H_1 |j\rangle=\langle i|H_{\rm eff}-H_0 |j\rangle=
  a^{\rm cl}_{i;j}-E_0^{\rm cl}\delta_{i,j}\ . 
\end{equation}
Since $H_1$ is a cluster additive, i.e., an extensive, operator, the
coefficients $t_{i;j}$ 
can be calculated for the infinite system on finite clusters up to
some finite order. This is the reason why we dropped the superscript cl from 
$t_{i;j}$. The cluster ground state energy $E_0^{\rm cl}$
must be calculated on the {\it same} cluster as the ``raw'' hopping
coefficients $a^{\rm cl}_{i;j}$.

For each order of the
coefficient $t_{i;j}$ there exists a 
minimum cluster which must contain the two rungs $i$ and
$j$. To classify the size of the minimum cluster we study how far
the triplon motion extends in a given 
order $l$. Only processes, which take
place on {\it linked} clusters of active bonds (see previous section),
contribute to the extensive thermodynamic hopping coefficients
$t_{i;j}$. The minimum cluster must be a linked cluster, which 
contains the rungs $i$ and $j$. 

\begin{figure}[h]
  \begin{center}
    \includegraphics[width=8.6cm]{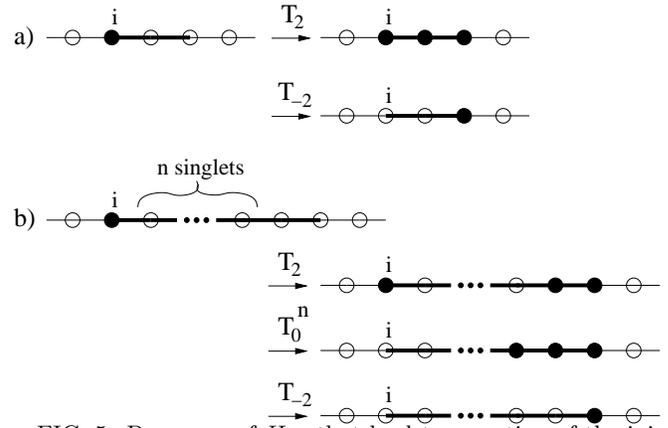}
    \caption{\label{fig_triphop_1} Processes of $H_{\rm eff}$ that
      lead to a motion of the initial triplon on rung $i$. Active
      bonds are depicted by thick lines. All processes that contribute
      to thermodynamic extensive hopping coefficients take place on linked
      clusters of active bonds. Part a) shows a second
      order process moving the initial triplon by two rungs. Part b)
      is a process of order $n+2$ moving the triplon by $n+2$ rungs.}
  \end{center}
\end{figure}

The action of a single $T_0$ operator (first order process) on
$|i\rangle$ is to shift the triplon by one as can be readily seen from
Tab.~\ref{matrix_tabI}. Somewhat more intricate is the case of the
operator $T_2$ acting on $|i\rangle$. In any operator-product
$T(\underline{m})$ an operator $T_2$ is always  accompanied by a
destruction operator $T_{-2}$. The operator $T_2$ creates two triplons
on  neighbouring sites (triplon-pair) if both of them are occupied by
singlets. Suppose that $T_2$ is immediately followed by the $T_{-2}$
operator. Then  there can be a hopping of the initial triplon by two
rungs, if the triplon-pair was created in the immediate vicinity of
the triplon at site $i$ to produce a three-triplon state. The
situation is depicted in Fig.~\ref{fig_triphop_1}a. This is a second
order process. It moves the triplon by two rungs.
We could go on like this ($\ldots T_{-2}T_2T_{-2}T_2$) or we could
start to build up a linked chain by iterative application of $T_2$
operators, say, to the right of the triplon at site $i$ and then destroy the
chain from the left (e.g. $T_{-2}T_{-2}T_2T_2$). All these processes
lead to a maximum motion of 
the initial triplon by $l$ rungs in $l^{\rm th}$ order.

The creation of a triplon-pair
{\it not} connected to the initial triplon on site $i$ does not lead to any
motion of the latter unless there is a sufficient number of $T_0$
operators moving the triplon at site $i$ towards the isolated triplon-pair
until they form a state with three adjacent triplons as depicted in
Fig.~\ref{fig_triphop_1}b. This also
leads to a maximum motion of the initial triplon by $l$ rungs in
$l^{\rm th}$ order. 

\begin{figure}[h]
  \begin{center}
    \includegraphics[width=6.0cm]{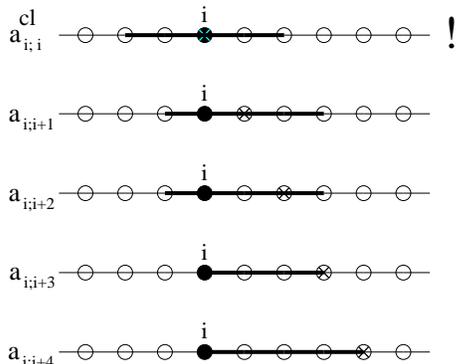}
    \caption{\label{fig_triphop_2} All possible hopping coefficients
      that can be calculated in 4$^{\rm th}$ order. Again, active
      bonds are depicted by thick lines. All processes that have to be
      considered take place on linked clusters. The initial (final) triplon
      positions are depicted by a filled circle (cross). They are
      contained in the minimum cluster (cl), which is defined by all active
      bonds for each coefficient. The exclamation mark next to the
      $a^{\rm cl}_{i;i}$ cluster 
      is to remind us that we have to subtract the cluster energy
      $E_0^{\rm cl}$ to get the cluster independent hopping coefficient
      $t_{i;i}=t_0$, c.f. Eq.~(\ref{hopp_amp_mod})}  
  \end{center}
\end{figure}

All possible combinations of the $T_2$, $T_0$ and
$T_{-2}$ operators that can appear in a $T(\underline{m})$-product of
$H_{\rm eff}$ can now be viewed as a product of the processes
discussed. So we conclude
\begin{eqnarray}
  \label{1_hopp_fazit}\nonumber
  &&\mbox{maximum motion of one triplon under the action}\\
&&\mbox{of }H_{\rm
  eff}\mbox{ in }l^{\rm th}\mbox{ order}=l\mbox{ sites .}
\end{eqnarray}
Therefore, the minimum cluster to calculate the hopping coefficient
$t_{i;j}$ in order $l$ in the thermodynamic limit must contain the two
rungs $i$ and $j$, which must not be further apart than $l$
rungs. Additionally the minimum cluster must contain all $l$ bonds
that can be activated in all processes moving the triplon from rung
$i$ to rung $j$. Fig.~\ref{fig_triphop_2} illustrates the situation
for all coefficients that can be calculated in fourth order. 

Because of translational invariance of the original underlying model
we can choose a suitable origin on 
each minimum cluster and it suffices to use a single label
for the hopping coefficients, i.e., $t_{i;j}=:t_{i-j}=t_d$. Additionally, one
has $t_d=t_{-d}$ due to inversion symmetry. Note that these relations
follow only for the thermodynamic hopping coefficients. In general, the
cluster specific coefficients $(t,a)^{\rm cl}$ have lower
symmetries. Following the argument above we calculate all
thermodynamic hopping coefficients in $14^{\rm th}$ order
($t_0,t_1,\ldots t_{14}$) on an {\it open} cluster of 15 rungs. 

From the thermodynamic cluster-independent hopping coefficients we 
construct the one-triplon energies. We define the Fourier-transformed
states 
\begin{equation}
  \label{fourier_states_1}
  |k\rangle=\frac{1}{\sqrt{N}}\sum_j e^{-ikj}|j\rangle\ ,
\end{equation}
where $j$ counts the rungs and $N$ is the total number of
rungs. Calculating the action of $H_1$ on these states
yields 
\begin{mathletters}
\begin{eqnarray}
  \label{1-trip_energ}
  H_1|k\rangle&=&\frac{1}{\sqrt{N}}
  \sum_{j,d=-l_{\rm max}}^{l_{\rm max}}   
  e^{-ikj}t_d|j+d\rangle\\
  &=&\frac{1}{\sqrt{N}}\sum_{j,d=-l_{\rm max}}^{l_{\rm max}}
  e^{-ik(j-d)}t_d|j\rangle\\
  &=&\sum_{d=-l_{\rm max}}^{l_{\rm max}}e^{ikd}t_d\underbrace{\frac{1}{\sqrt{N}}\sum_je^{-ikj}|j\rangle}_{|k\rangle}\ .
\end{eqnarray}
\end{mathletters}
Making use of the inversion symmetry $t_d=t_{-d}$ yields the real one-triplon
dispersion
\begin{equation}
  \label{1-energ}
  \omega(k;x)=\langle k| H_1(x)|k\rangle=t_0+2\sum_{d=1}^{l_{\rm max}}
  t_d\cos(dk)\ ,  
\end{equation}
where the maximum order $l_{\rm max}$ is 14 in our case. 

Again, the
hopping coefficients $a_{i;j}$ and the corresponding cluster ground state
energy $E_0$ and therefore the cluster-independent coefficients $t_d$
are calculated by implementing the action of $H_{\rm eff}$ on the
states $|i\rangle$ on a
computer (details in Ref~\cite{knett99a}). For fixed one-triplon momentum $k$
the one-triplon dispersion is a 14$^{\rm th}$ order polynomial in $x$ with
real coefficients. Thereby we retrieve and extend the numerical
13$^{\rm th}$ order result in Ref.~\cite{weiho98a}.

\begin{figure}[h]
  \begin{center}
    \includegraphics[width=8.6cm]{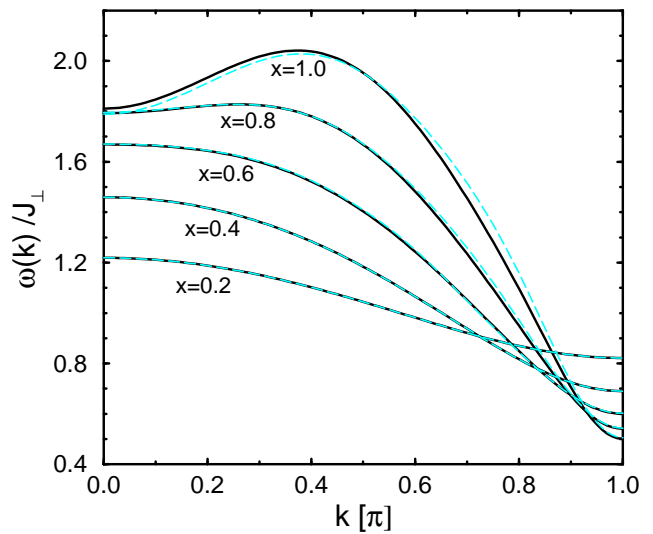}
    \caption{\label{disp_vergl} One-triplon dispersion for various
      $x$-values as indicated. The grey dashed lines correspond to
      plain series results ($x=$0.2, 0.4, 0.6) or to optimized series
       results ($x=$0.8, 1.0). The solid curves depict our most
      reliable results obtained by the novel extrapolation scheme
      explained in Ref.~\protect\cite{schmi02}.}
  \end{center}
\end{figure}

In Fig.~\ref{disp_vergl} the one-triplon dispersion is displayed for five
different $x$-values. 
For $x=$0.2, 0.4 and 0.6 the grey dashed curves represent the results
obtained by using the plain series results for the hopping coefficients
$t_d$. The grey dashed curves for $x=$0.8 and 1.0 result from
using the optimized hopping coefficients according to the optimized
perturbation theory explained in Sect.~\ref{OPT_section} (parameter
choice: $\alpha_{\rm opt}=2.9x$). These results are compared to the
most reliable extrapolations depicted as solid lines. The latter are
obtained by using the novel extrapolation scheme based on a re-expansion
of the original series results (polynomials in $x$) in terms of a
suitable internal variable $p(x)$~\cite{schmi02}.

\subsection{Two-Triplon Interaction: $H_2$}
We define the states $|i,j\rangle$, denoting the eigen state of
$H_{\perp}$ with triplon 1 on rung $i$, triplon 2 on rung
$j$ and singlets on all other rungs. Two triplons together
can form an $S=0$ singlet, an $S=1$ triplet or an $S=2$
quintuplet bound state. Tab.~\ref{tot_spin} summarizes these nine states
sorted by their total spin $S$ and magnetic quantum number 
$m$. 

By construction $H_{\rm eff}$ conserves the total spin $S$ and the magnetic
quantum number $m$. Therefore it
is convenient to work 
in the basis given in Tab.~\ref{tot_spin}. This table defines the
states $|i,j\rangle^{S,m}$ by the linear combinations in the third column.
\begin{table}[h]
\begin{tabular}{|c|c|c|}
\phantom{m}$S$\phantom{m} &  $m$ &  \\
\hline\hline
 2&  2 & $|t^1,t^1\rangle$\\
 2&  1 & $1/\sqrt{2}(|t^0,t^1\rangle + |t^{1},t^0\rangle)$\\
 2&  0 & $1/\sqrt{6}(|t^{-1},t^{1}\rangle+2|t^{0},t^{0}\rangle+|t^1,t^{-1}\rangle)$\\
 2& -1 & $1/\sqrt{2}(|t^{-1},t^{0}\rangle+|t^0,t^{-1}\rangle)$\\
 2& -2 & $|t^{-1},t^{-1}\rangle$\\
\hline
 1&  1 & $1/\sqrt{2}(|t^1,t^0\rangle-|t^0,t^1\rangle)$\\
 1&  0 & $1/\sqrt{2}(|t^1,t^{-1}\rangle-|t^{-1},t^1\rangle)$\\
 1& -1 & $1/\sqrt{2}(|t^0,t^{-1}\rangle-|t^{-1},t^0)$\\
\hline
 0&  0 & $1/\sqrt{3}(|t^0,t^0\rangle-|t^1,t^{-1}\rangle-|t^{-1},t^1\rangle)$\\
\end{tabular}
\caption{\label{tot_spin} The nine states two triplons can
  form, combined to states with given quantum numbers $S$ and $m$.}  
\end{table}

Again, due to triplon conservation the action of 
$H_{\rm eff}$ on the state $|i,j\rangle$ is to shift the triplons to
rung $i'$ and rung $j'$ conserving also $S$ and $m$. Nothing else is
possible. In analogy to 
Eq.~(\ref{hopp_amp}) of the preceding section we define the interaction
coefficients 
\begin{equation}
  \label{2-hopp-amp}
  a^{S,{\rm cl}}_{ij;kl}(x)=\hspace*{1mm} ^S\hspace*{-0.2mm}\langle i,j|H_{\rm eff}(x)|k,l\rangle^S\ .
\end{equation}
The coefficients depend on the total spin $S$ but not on the magnetic
quantum number $m$. Hence the $m$-index is dropped here and in the
following.  

The exchange parity is determined by the total spin $S$
\begin{equation}
  \label{exch_par}
  |i,j\rangle^S=(-1)^S|j,i\rangle^S\ .
\end{equation}
This means that we can restrict the description to those states
$|i,j\rangle$ for which $i<j$. 

Making use of the above the irreducible two-triplon interaction
coefficients $t^S_{ij;kl}$ follow from 
\begin{align}
  \nonumber 
  t_{ij;kl}^S=&\ ^S\!\langle i,j|H_2|k,l\rangle^S =\  ^S\!\langle i,j|H_{\rm
  eff}-H_1-H_0|k,l\rangle^S\\ 
  \nonumber
  = &a^{S,{\rm cl}}_{ij;kl}-E_0^{\rm cl}\delta_{i,k}
  \delta_{j;l}\\\label{S_conservation} 
  -&t^{{\rm cl}}_{i;k}\delta_{j,l}-t^{{\rm cl}}_{j;l}\delta_{i,k}-t^{{\rm
  cl}}_{i;l}\delta_{j,k} (-1)^S
  -t^{{\rm cl}}_{j;k}\delta_{i,l}(-1)^S\ . 
\end{align}
analogous to Eq.~(9) in Ref.~\cite{knett03a}. 
Again, $E_0^{\rm cl}$ and the one-triplon hopping coefficients
$t_{i;j}^{\rm cl}$ must be calculated on the {\it same} cluster as
the ``raw'' two-triplon coefficients $a_{ij;kl}^{\rm cl}$.
The cluster hopping coefficients $t^{{\rm cl}}_{j;l}$ are needed only
in the intermediate steps of the calculation of the irreducible
interaction coefficients.

There will be no $t_{ii;kl}$ or $t_{ij;kk}$ since it is not possible
to have two triplons on one rung at the same time. This constraint
can be viewed as a hardcore repulsion interaction. 

\begin{figure}[h]
  \begin{center}
    \includegraphics[width=6.6cm]{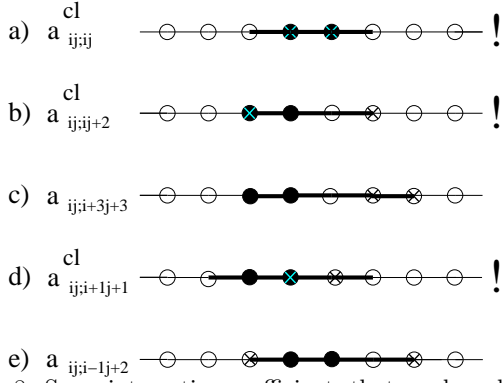}
    \caption{\label{2trip_hopp_fig1} Some interaction coefficients that can
      be calculated in $4^{\rm th}$ order. Initial (final) triplon
      pairs are denoted by full circles (crosses). The thick solid
      lines are active bonds between rungs. Exclamation marks indicate
      that one has to subtract one- or zero-triplon terms according to
      Eq.~(\ref{S_conservation}) to obtain the extensive
      thermodynamic interaction coefficients.}
      
  \end{center}
\end{figure}

The construction of the minimum cluster needed to calculate the
$t_{ij; kl}$ in the thermodynamic limit follows the same line of
argumentation as in the one-particle section. Generally, the cluster 
must be large enough to encompass all possible processes in 
order $l$. The minimum cluster has to include all linked bonds that
can be activated in any possible interaction process of length $l$ which
leads to state 
$|i',j'\rangle$ if one starts with state $|i,j\rangle$. Obviously the
rungs $i$, $j$, $i'$ and $j'$ must be contained in the minimum
cluster and they must be connected by active
bonds. Fig.~\ref{2trip_hopp_fig1} shows some interaction coefficients with
fixed initial configuration (adjacent triplons) and their associated 
minimum clusters in 4$^{\rm th}$ order. 
All interaction
coefficients of order $l$ can be calculated on a cluster containing
$l+1$ rungs.

For particular systems there may be symmetries, e.g., 
spin rotation invariance, 
or other particularities, e.g., nearest-neighbour coupling only, 
which prevent certain processes from generating non-vanishing coefficients. 
Indeed, we found that the spin ladder with nearest-neighbour coupling in order 
$l$ induces only interaction coefficients of a range which can be determined
by distributing $l$ hops among the two triplons.
For instance, in 4$^{\rm th}$ order the triplon at rung $i$  
may hop to rung $i-1$ and the triplon at rung $j$ hops to rung
$j+3$. Another possible process might be that the triplon at rung $i$
stays at this rung while the triplon at rung $j$ hops to rung $j+4$ or
maybe only to rung $j+3$ and so on.
This particularity implies for instance that the process
shown in Fig.~\ref{2trip_hopp_fig1}c vanishes for the spin ladder
system.

Due to the translational invariance of the ladder system the momentum $k$
is a good quantum number in the one-triplon sector and the diagonal 
matrix elements of the Fourier
transformed states $|k\rangle$ are the eigen energies
$\omega(k)$. With two triplons present only the {\it total} momentum
$K$ is a good quantum number. The relative momentum $q$ is not
conserved and generally leads to the formation of a two-triplon
continuum. 

To make use of the conserved total momentum we turn to a 
new basis. As a first step we use center-of-mass coordinates,
i.e., $|i,j\rangle^S\rightarrow |r,r+d\rangle^S=(-1)^S|r+d,r\rangle^S$,
with $r=i$ and $d=j-i$. The restriction $i<j$ (see
text below Eq.~(\ref{exch_par})) translates to $d>0$. We choose a
suitable origin, say $k=0$, and rename 
\begin{equation}
  \label{t_drd-prim}
  t_{d;r,d'}\equiv \langle
  r,r+d'|H_2|0,d\rangle =\langle i,j|H_2|k,l\rangle =  t_{ij;kl}\ ,
\end{equation}
with $d=l$, $r=i$ and $r+d'=j$. From Eq.~(\ref{S_conservation}) we
then obtain 
\begin{align}
  \nonumber
  t^{S}_{d;r,d'}&=a^{S,{\rm cl}}_{d;r,d'}-E_0^{\rm
    cl}\delta_{0,r}\delta_{d,r+d'}-t^{\rm cl}_{r;0}\delta_{d,r+d'} -t^{\rm
    cl}_{d;r+d'}\delta_{0,r}\\\label{t_drd-prim_fin}
  &-t^{\rm cl}_{0;r+d'}\delta_{d,r}(-1)^S -t^{\rm
    cl}_{d;r}\delta_{0,r+d'}(-1)^S 
\end{align}
in the new basis. This is equivalent to the equations
emerging from considering the special cases
\begin{mathletters}
\label{WW-pure}
\begin{eqnarray}
t^S_{d;0,d'}&=&a^{S,{\rm cl}}_{d;0,d'}-t^{\rm
  cl}_{d'-d}-\delta_{d,d'}(t^{\rm 
  cl}_0+E^{\rm cl})\\ 
t^S_{d;d-d',d'}&=&a^{S,{\rm cl}}_{d;d-d',d'}-t^{\rm
  cl}_{d-d'}-\delta_{d,d'}(t^{\rm cl}_0+E^{\rm cl})\\ 
t^S_{d;-d',d'}&=&a^{S,{\rm cl}}_{d;-d',d'}-t^{\rm cl}_{-d-d'}(-1)^S\\
t^S_{d;d,d'}&=&a^{S,{\rm cl}}_{d;d,d'}-t^{\rm cl}_{d+d'}(-1)^S\ .
\end{eqnarray}
\end{mathletters}
Otherwise the interaction coefficients $t^S_{d;r,d'}$ and $a^S_{d;r,d'}$ are
identical.

As a second step the states $|r,r+d\rangle^S$ are Fourier transformed
with respect to the center-of-mass variable $(r+d/2)$ 
\begin{align}
  \nonumber
  |K,d\rangle^S &:=\ \ \ \ \ \ \ \frac{1}{\sqrt{N}}\sum_r
  e^{iK(r+d/2)}|r,r+d\rangle^S \\\nonumber
  & =(-1)^S\frac{1}{\sqrt{N}}\sum_r
  e^{iK(r+d/2)}|r+d,r\rangle^S \\\nonumber
  &\!\!\!\!\! \overset{r\rightarrow r+d}{=}(-1)^S\frac{1}{\sqrt{N}}\sum_r
  e^{iK(r-d/2)}|r,r-d\rangle^S \\\label{basis}
  & =(-1)^S|K,-d\rangle^S\ ,
\end{align}
where $K$ is the conserved total momentum in the Brillouin
zone and $N$ is the number of rungs. For fixed $K$ and $S$ the relative
distance $d>0$ between two triplons is the only remaining quantum
number one has to keep track of.

To obtain the complete two-triplon excitation energies we have to
calculate the action of 
\begin{equation}
  \label{omega_2}
  H_{\rm eff}-H_0=H_1+H_2\ 
\end{equation}
on the two-triplon states $|K,d\rangle$. The two addends on the right
hand site are considered separately in the following. 

\begin{figure}[h]
  \begin{center}
    \includegraphics[width=7.8cm]{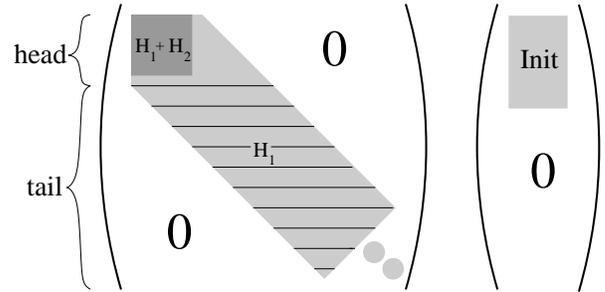}
    \caption{\label{T_W_SV_matrix} The left part of the figure
      schematically shows
      the matrix representation of $H_1$ and $H_2$ in the two-triplon
      $\{|K,d\rangle\}$ basis~(\ref{basis}). The matrix $H_1$ has
      elements in the whole grey area, while $H_2$ has elements in the
      dark grey area only. We calculated the elements of $H_2$ up to
      order 13 and those of $H_1$ up to order 14, so that $H_2$ is a
      finite $13\times 13$ matrix and $H_1$ a semi-infinite
      band matrix, whose width is 28, see Eqs.~(\ref{T}) and
      (\ref{W_2}) for further information. The sum of $H_1$ and $H_2$
      represents $H_{\rm eff}$ in the two-triplon sector to the given
      orders. The right part shows the
      initial vector $|\text{Init}\rangle={\mathcal O}_{\rm eff}|0\rangle$ as
      calculated in Sect.~\ref{O_transformation} for the two-triplon 
      sector. Since we calculated ${\mathcal O}_{\rm eff}$ up to
      order 10,  $|\text{Init}\rangle$ is a vector of dimension 10 in the
      $\{|K,d\rangle\}$ basis. The Green function ${\mathcal G}$
      (Eq.~(\ref{green_1})) 
      is calculated by tridiagonalization, more
      information in Sect.~\ref{sec_G_function}. For $K$ and $x$
      fixed, the elements of the matrix and the vector reduce to
      real numbers.} 
  \end{center}
\end{figure}

The operator $H_1$ can move one
of the two triplons at maximum. A short calculation yields
\begin{eqnarray}
  \nonumber
  &&H_1|K,d\rangle^S=\\\nonumber
 &&\frac{1}{\sqrt{N}}\sum_re^{iK(r+d/2)}
  \!\!\!\!\sum_{n=-l_{\rm max} \atop n\not= d}^{l_{\rm max}}\!\!\!
  t_n(|r+n,r+d\rangle^S+|r,r+d-n\rangle^S)\\\nonumber
  &&=\sum_{n=-l_{\rm max}\atop n\not= d}^{l_{\rm
      max}}t_n(e^{iKn/2}+e^{-iKn/2})\times\\\nonumber
  &&\ \ \ \ \times\underbrace{\frac{1}{\sqrt{N}} \sum_r
  e^{iK(r+(d-n)/2)} |r,r+d-n\rangle^S}_{|K,d-n\rangle}\\\label{T}
  &&=2\sum_{n=-l_{\rm max} \atop n\not= d}^{l_{\rm max}} t_{n}\cos\left(K\frac{n}{2}\right)\left[{\rm
  sgn}(d-n)\right]^S|K,|d-n|\rangle^S \ .
\end{eqnarray}
Here we used the previously calculated matrix-elements
$t_n=t_{-n}$ (inversion symmetry), which have been calculated to
$l_{\rm max}=14$ (cf.  
preceding subsection). Since we restricted $d>0$ the sgn-function
enters the result by  Eq.~(\ref{basis}). 
For fixed $K$, $H_1$ now appears as a semi-infinite band
matrix in the remaining quantum number $d$. Independent of the size of
the initial distance $d>0$ between the two triplons, $H_1$ will
produce states where the distances between the triplons are
incremented or decremented by 14 rungs ($l_{\rm max}=14$) at maximum. If
the initial 
distance $d$ is larger than 14, $H_1$ continues to produce the same
matrix elements on and on for all $d>14$, i.e., the matrix representing
$H_1$ in the chosen basis for fixed $K$ is semi-infinite with a
repeated pattern in the tail. The head of $H_1$, i.e., the 14$\times$14
block between states with $d\le14$, contains matrix elements with a
somewhat more complicated structure. Here the matrix element between the
starting distance $d$ and the final distance $d'$ is a sum of the
direct process $d\rightarrow d'$, where one of the triplons has hopped
$n$ rungs to the right ($n>0$) or to the left ($n<0$) with $d-n=d'>0$,
and the indirect process with 
$d-n=-d'<0$. The situation is sketched in
Fig.~\ref{T_W_SV_matrix}. The matrix $H_1$ comprises the full
thermodynamic one-triplon dynamics in the two-triplon sector for the
given order $l_{\rm max}=14$.

The situation is more complex for $H_2$. In a first step
we find 
\begin{align}
  \nonumber
  & H_2|K,d\rangle^S = \\\nonumber
  &\frac{1}{\sqrt{N}}\sum_re^{iK(r+d/2)}
 \!\!\!\!\!\!\!\!\sum_{\max\{n+d',d-n\}\atop\le l_{\rm
     max}}\!\!\!\!\!\!\!\! t_{d;n,d'}|r+n,r+n+d'\rangle^S=\\\label{W_1}
 & \!\!\!\!\!\!\!\!\sum_{\max\{n+d',d-n\}\atop\le l_{\rm
     max}}\!\!\!\!\!\!\!\! t_{d;n,d'} e^{iK(-n+(d-d')/2)}|K,d'\rangle^S\ , 
\end{align}
with the two integers $n\in {\mathbb Z}$ and $d'\in{\mathbb N}$ as
summation indices. The 
positive distances $d$ and $d'$ must be smaller or equal to $l_{\rm max}$,
since a maximum of $l_{\rm max}$ linked bonds can be produced in this
order and all four triplons sites (the two initial sites and the two
final sites) must be contained in the resulting linked cluster. The
$t_{d;n,d'}$ are the matrix elements of $H_2$ defined in
Eq.~(\ref{t_drd-prim_fin}). The last equality follows from
substituting the summation-index $r\rightarrow r+n$.

To  simplify the expression further inversion symmetry is used. We have
\begin{equation}
  t_{d;r',d'}=
  \langle r'',r''+d'|H_2|r,r+d\rangle\ ,
\end{equation}
with $r'=r''-r$.
The thermodynamic interaction coefficient $t_{d;r',d'}$ is
associated with a fixed constellation of initial and final triplon
pairs. We define a configuration CON by the set of four positions
given by these two pairs ${\rm
  CON}=\{r,r+d,r'',r''+d'\}$. Let $s$ denote
the middle of this configuration $s=(\max({\rm CON})-\min({\rm
  CON}))/2$. Reflecting a configuration about $s$ and  
interchanging the triplon positions in both initial and final triplon
pairs  gives 
\begin{align}
  \nonumber t_{d;r',d'}&=\langle
  r'',r''+d'|H_2|r,r+d\rangle\\\nonumber
  =&\langle
  2s-r''-d',2s-r''|H_2|2s-r-d,2s-r\rangle\\ &=
  t_{d;d-d'-r',d'}\ .
\end{align}
Possible minus signs cancel since they appear twice.
We can now split the sum over $n$ in Eq.~(\ref{W_1}) in three parts,
$n>(d-d')/2$, $n<(d-d')/2$ and $n=(d-d')/2$. The second sum is indexed
back to $n>(d-d')/2$ by making use of
$\sum_{n<j}a_n=\sum_{n>j}a_{2j-n}$ where $j:=(d-d')/2$
\begin{align}
  \nonumber
  H_2|K,d\rangle^S&= \hspace*{-10mm}\sum_{\max\{n+d',d-n\}\le l_{\rm
  max} \atop 
  n>(d-d')/2\in {\mathbb Z}}\hspace*{-10mm} \left[ t_{d;n,d'}
  e^{iK(-n+(d-d')/2)}|K,d'\rangle^S \right .\\\nonumber
  &\hspace*{10mm}\left . +t_{d;d-d'-n,d'}e^{iK(n-(d-d')/2)}
  |K,d'\rangle^S\right]\\\nonumber
  &\hspace*{4mm}+\hspace*{-10mm} \sum_{\max\{n+d',d-n\}\le l_{\rm max} \atop
  n=(d-d')/2 \in {\mathbb Z}} \hspace*{-10mm}
  t_{d;(d-d')/2,d'}|K,d'\rangle^S\\\nonumber
  &=2\hspace*{-10mm}\sum_{\max\{n+d',d-n\}\le l_{\rm
  max} \atop 
  n>(d-d')/2 \in {\mathbb Z}}\hspace*{-10mm}
  t_{d;n,d'}\cos[K(n-(d-d')/2)]|K,d'\rangle^S\\\label{W_2} 
  &\hspace*{4mm}+\hspace*{-10mm} \sum_{\max\{n+d',d-n\}\le l_{\rm max} \atop
  n=(d-d')/2\in {\mathbb Z}} \hspace*{-10mm}
  t_{d;(d-d')/2,d'}|K,d'\rangle^S\ . 
\end{align}

In contrast to $H_1$ the matrix representing $H_2$ is of finite
dimension due to the finite range of the contributing processes (finite
maximum order) expressed by the restrictions of the sums appearing in
Eq.~(\ref{W_2}). In our case
the $t_{d;r,d'}$ have been calculated up to $l_{\rm max}=13$ giving
rise to a $13\times 13$ matrix in the distance $d$ for fixed
$K$. Fig.~\ref{T_W_SV_matrix} sketches the situation.

Finally, the sum of the two matrixes $H_1$ and $H_2$ with respect to
basis~(\ref{basis}) comprises the complete two-triplon dynamics.

\begin{figure}[h]
  \begin{center}
    \includegraphics[width=8.4cm]{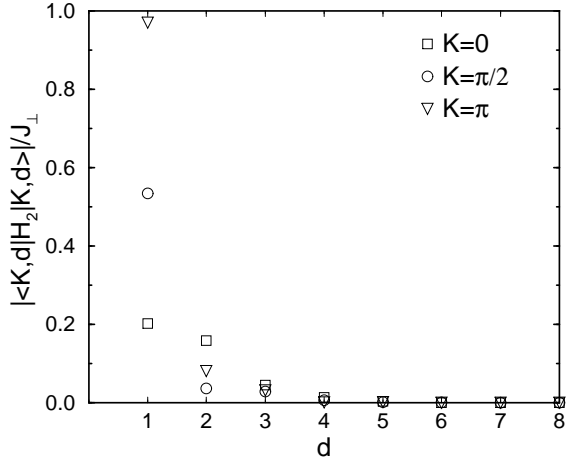}
    \caption{\label{H_2_dropoff} Expectation values of some diagonal
      elements of $H_2$ (pure two-triplon interaction) for the ladder
      system in the
      $\{|K,d\rangle\}$ basis for various
      $K$-values and $x$ set to 0.6 as function of the remaining
      quantum number $d$. Clearly, the irreducible two-triplon
      interaction coefficients drop off rapidly with increasing relative
      distance $d$
      between the two triplons. Non-diagonal elements behave in a
      similar fashion. At larger distances two triplons are
      asymptotically free.} 
  \end{center}
\end{figure}

The above approach is well justified. At large distances the
two-triplon dynamics is governed by independent one-triplon hopping. At
smaller distances an additional two-particle interaction occurs given by
$t_{d;r,d'}$ connecting the state $|r,r+d'\rangle$ with state
$|0,d\rangle$. The sum $H_1+H_2$ gives the combined effect of
one-triplon  hopping and two-triplon
interaction. Fig.~\ref{H_2_dropoff} shows that the interaction
coefficients $\langle K,d|H_2|K,d\rangle$ for the ladder system indeed
drop off rapidly for larger distances.  


Taking the perturbation expansion up to order $l_{\rm max}$ allows to
calculate the irreducible two-particle interaction up to a distance $l$
between the 
two particles correctly within order $l_{\rm max}$. No 
processes involving larger distances appear. But the part of the two-particle
sector that can be described by one-particle dynamics alone is
taken into account for {\it all} distances between the two particles
and describes hopping processes of range $\le l_{\rm max}$ correctly
within order $l_{\rm max}$. 

At $K=\pi$, the smallest eigen value of $\tilde{H_1}+H_2$, where
$\tilde{H_1}$ is the upper left $13\times 13$ sub-matrix of $H_1$, can be
extracted as a 13$^{\rm th}$ order polynomial in $x$ and identified as
the (lowest) bound state. This is possible because at $K=\pi$, the relative
motion of the two triplons is of order $x^2$ while
 the interaction enters in order $x$. Hence the interaction dominates
 over the kinetics for $x\to 0$ so that a local bound pair is the simple
eigen state for vanishing $x$.
Our results extend the results by Zheng et
al.~\cite{weiho00a} for $S=1$ from 12$^{\rm th}$ to 13$^{\rm th}$
order, and for $S=0$ from 7$^{\rm th}$ to 13$^{\rm th}$ order. The
polynomials will be made available on our web pages~\cite{homepages}.

\section{Transformation of the observables}
\label{O_transformation}
\subsection{General Aspects}
The continuous unitary transformation of observables has been
explained in detail in Ref.~\cite{knett03a}. Here we briefly review
the most important general aspects before we describe the procedure
for the spin ladder in detail.

Using the {\it same} transformation as for the Hamiltonian we derive
a series expansion (similar to Eq.~(\ref{H_eff})) for the effective
observable ${\mathcal O}_{\rm eff}$ onto which a given initial
observable ${\mathcal O}$ is mapped by the perturbative CUT procedure
\begin{equation}
  \label{O_eff_1}
  {\mathcal O_{\rm eff}}(x) = \sum_{k=0}^{\infty}x^k\sum_{i=1}^{k+1}
  \sum_{|\underline{m}|=k}\tilde{C}(\underline{m};i){\mathcal
  O}(\underline{m};i)\ ,
\end{equation}
where 
\begin{equation}
\label{O_ansatz_II}
{\mathcal O}(\underline{m};i):=T_{m_1}\cdots T_{m_{i-1}}{\mathcal O}
  T_{m_i}\cdots T_{m_k}\ .
\end{equation}
The operators $T_i$ are the same as in the Hamilton
transformation. The coefficients $\tilde{C}(\underline{m};i)$ can
again be calculated on a computer. They are fractions of
integers~\cite{knett03a}.

The effective observables are described by weighted virtual
excitation processes $T(\underline{m})$ interrupted by 
processes induced by the observable as given in 
Eq.~(\ref{O_ansatz_II}). Sometimes it is convenient
to seek for a decomposition of ${\mathcal O}$ with respect to its action
on the number of particles
\begin{equation}
  \label{O_decomp_1}
  {\mathcal O}=\sum_{n=-N}^N T'_n\ ,
\end{equation}
where $T'_n$ creates $n$ particles or destroys them if $n<0$.

An important point is that ${\mathcal O}_{\rm
  eff}$ is {\it not} an energy quanta conserving quantity, i.e., it
does not conserve the number of triplons in the spin ladder
system. This is 
formally expressed by the fact that the sum over $\underline{m}$ is
not restricted to $M(\underline{m})=0$, so that ${\mathcal O}_{\rm
  eff}$ can add or subtract an arbitrary number of particles.

The effective operators ${\mathcal O}_{\rm eff}$ can be decomposed in
a sum of cluster-additive operators ${\mathcal O}_{p,n}$, for which
the linked cluster theorem can be used 
\begin{equation}
  \label{O_decomp}
  {\mathcal O}_{\rm eff}=\sum_{n=0}^{\infty}\sum_{p\ge -n}
  {\mathcal O}_{p,n}\ .
\end{equation}
Here $p$ indicates how many particles are created
($p\ge 0$) or destroyed ($p<0$) by ${\mathcal O}_{p,n}$. The subindex
$n\ge 0$ indicates the minimum number of particles that must be
present for ${\mathcal O}_{p,n}$ to have a non zero action. The action
of the operator ${\mathcal O}_{p,n}$ on a state containing less than
$n$ particles is zero. Further definitions and details concerning
the operators ${\mathcal O}_{p,n}$ can be found in
Ref.~\cite{knett03a}. 

Focusing on $T=0$ experiments in
the following, we treat only the operators
${\mathcal O}_{p\ge 0,0}$. Their interpretation is particularly
simple. The effective observable ${\mathcal O}_{\rm eff}$
acting on the $T=0$ state, i.e., the ground state or excitation vacuum,
respectively, decomposes in a
sum of the operators ${\mathcal O}_{p\ge 0,0}$, each injecting
$p=0,1,2\ldots$ triplons into the system. 
In Ref.~\cite{knett03a} we showed that the ${\mathcal
  O}_{p\ge 0,0}$ can be directly calculated from the action of
${\mathcal O}_{\rm eff}$ on $|0\rangle$ on minimum finite
clusters. No extra terms have to be subtracted to obtain thermodynamic
results. The calculations can again be implemented on a computer in
analogy to what was done for the Hamiltonian.

To be more specific let ${\mathcal O}$ be a {\em locally} acting
observable injecting triplons at a specific site $r$ of
the ladder. Then the effective observable reads
\begin{align}
  \nonumber
  &{\mathcal O}_{\rm eff}(r)|0\rangle = \sum_{p\ge 0} {\mathcal
  O}_{p,0}(r)|0\rangle\\\nonumber 
  &=c|0\rangle + 
  \sum_{n=-l_{\rm max}}^{l_{\rm max}}c_{n}|r+n\rangle
  +\\\label{O_i123_def} 
  &\hspace*{3mm}+\sum_{n,n'\atop |n|+|n'|\le l_{\rm
  max}}c_{n,n'}|r+n,r+n'\rangle + \cdots \ . 
\end{align}
The restriction $|n|+|n'|\le l_{\rm max}$ for the third sum reflects the
fact that the two triplons, after being injected, cannot undergo more
rung-to-rung hops in total than the maximum order $l_{\rm
  max}$. Therefore, the maximum distance $p=|n-n'|$ occurring
is $l$ in order $l$ for the spin ladder system.

Once the coefficients $c$ are calculated the
spectral weights $I_{N}$ are accessible, which are contained in the
different particle-sectors characterized by the number ${N}$ of
particles injected 
\begin{align}
\nonumber
  I_{N}&=\langle 0| {\mathcal O}_{-N,0}(r){\mathcal O}_{N,0}(r)
  |0\rangle\\\nonumber 
  &= \sum_{n_1,\ldots,n_{N}}|\langle r+n_1,\ldots,r+n_{N}|{\mathcal
    O}_{N,0}(r)|0\rangle|^2\\\label{tot_int}
  &= \sum_{n_1,\ldots,n_{N}}|c_{n_1,\ldots,n_{N}}|^2\ .
\end{align}
If the total weight $I_{\rm tot}$ of the operator is also known,
for instance via the sum rule $I_{\rm tot}=\langle 0|{\mathcal
  O}^2|0\rangle - \langle 0|{\mathcal O}|0\rangle ^2$, the {\it
  relative} weights of the individual particle sectors 
$I_N/I_{\rm tot}$ can be calculated. They serve as an important
criterion to judge the 
applicability of our approach. If most of the weight can be found in
sectors of low quasi-particle number and sectors of higher particle
number can be safely neglected the approach will work fine. The
chosen particles constitute a suitable basis to describe the system. 
This argument has been used by Schmidt and Uhrig~\cite{schmi03a} to
show, that the triplon is a well suited particle to describe the one-
dimensional spin chain. There basically all the spectral weight is
captured by one and two triplons.

So far {\it local} observables ${\mathcal O}(r)$ were
considered. A real experiment, however, couples to the system in a
global fashion. Due to translational invariance the injected particles
(here triplons)
have a total momentum $K$. Thus we define the {\it global}
observables in momentum space representations
\begin{align}
  \nonumber
  {\mathcal O}_{\rm eff}(K)|0\rangle&=\sum_{p\ge 0}{\mathcal
    O}_{p,n}(K)|0\rangle\\\label{O_K_1} 
  &=\sum_{p\ge 0}\frac{1}{\sqrt{N}}\sum_{r=1}^Ne^{iKr} {\mathcal
  O}_{p,0}(r)|0\rangle \ , 
\end{align}
where $N$ is the number of rungs in the system. Let us investigate the
one- and two-triplon sectors separately. In the one-triplon sector we
have (here $K$ is the one-triplon momentum $k$) 
\begin{align}
  \nonumber
  {\mathcal O}_{1,0}(k)|0\rangle&=\frac{1}{\sqrt{N}}
  \sum_{r} e^{ikr}\sum_{n}
  c_n|r+n\rangle\\\nonumber 
  &=\sum_nc_ne^{-ikn}
  \frac{1}{\sqrt{N}}\sum_re^{ikr}|r\rangle\\\label{O_K_2}
  &=\sum_nc_ne^{-ikn}|k\rangle \ .
\end{align}
We used the same definition for $|k\rangle$ as introduced in
Sect.~\ref{H_transformation}. Due to inversion symmetry $c_n=c_{-n}$
holds. Thus Eq.~(\ref{O_K_2}) simplifies to
\begin{align}
  \label{O_K_3}
  \langle k|{\mathcal O}_{1,0}(k)|0\rangle &=A_k=c_0+2\sum_{n} c_n\cos(kn)\ .
\end{align}
Somewhat more complex is the two-triplon sector 
\begin{align}
  \nonumber
  &{\mathcal O}_{2,0}(K)|0\rangle =\frac{1}{\sqrt{N}}\sum_{r=1}^N
  e^{iKr}\sum_{n,n'}c_{n,n'}|r+n,r+n'\rangle\\\nonumber
  &=\sum_{n,d}c_{n,n+d}e^{-iK(n+d/2)}\frac{1}{\sqrt{N}}\sum_r
  e^{iK(r+d/2)}|r,r+d\rangle\\\nonumber
  &=\sum_{n,d}c_{n,n+d}e^{-iK(n+d/2)}|K,d\rangle\\\label{O_K_4}
  &=\sum_dA_{K,d}|K,d\rangle\ ,
\end{align}
where we defined the relative distance $d=n'-n$ between the two
injected triplons. The definition of $|K,d\rangle$ is taken from
Sect.~\ref{H_transformation}. Again, inversion symmetry, here
$c_{n,n'}=(-1)^Sc_{-n,-n'}$, can be used
to obtain real results for the coefficients $A_{K,d}$. The variable
$S\in\{0,1,2\}$, which is a good quantum number, denotes the total
spin of the injected triplon pair.  

The action of 
${\mathcal O}_{\rm eff}$ from the ground state into the two-triplon
space produces the states $|K,d\rangle$ with $0<d\le l_{\rm max}$ in
order $l_{\rm max}$. Thus, for fixed $K$, 
the action of ${\mathcal O}_{\rm eff}$ may be visualized as
a vector in the remaining quantum number $d$ of which the first
$l_{\rm max}$ entries are the $A_{K,d}$ of
Eq.~(\ref{O_K_4}). All other entries are
zero. This vector, labeled initial vector $|\text{Init}\rangle$ for reasons
given in Sect.~\ref{sec_G_function}, is depicted in
Fig.~\ref{T_W_SV_matrix} together with the matrix representing $H_{\rm
  eff}$ for fixed $K$ in the two-triplon sector.

\subsection{Observables in the Spin Ladder}
We now turn to the evaluation of the observables of interest in the
ladder system. We calculated the $\tilde{C}(\underline{m};i)$ in
Eq.~(\ref{O_eff_1}) up to and including order $l_{\rm max}=10$ for the
problem under study. The four {\it local} operators considered are
\begin{mathletters}
  \label{ops}
  \begin{eqnarray}
    \label{o1}
    {\mathcal O}^{\rm I}(r) &=& {\mathbf S}_{1,r} {\mathbf
    S}_{2,r}={\mathcal T}^{\rm I}_0 \\
    \label{o2}
    {\mathcal O}^{\rm II}_l(r) &=& {\mathbf S}_{l,r}
      {\mathbf S}_{l,r+1}\\\nonumber
      &=&\frac{1}{4}\left( {\mathcal T}_{-2}+{\mathcal
      T}_0+{\mathcal T}_2+{\mathcal T}^{\rm II}_{-1}+{\mathcal
      T}^{\rm II}_1 \right)\\  
    \label{o3}
    {\mathcal O}^{\rm III}(r) &=& {\mathbf S}^{z}_{1,r}-{\mathbf
      S}^{z}_{2,r} = {\mathcal T}^{\rm III}_{-1}+ {\mathcal T}^{\rm
      III}_1\\ 
    \label{04}
    {\mathcal O}^{\rm IV}(r) &=& {\mathbf S}^{z}_{1,r}+{\mathbf
      S}^{z}_{2,r} = {\mathcal T}^{\rm IV}_0 \ ,
  \end{eqnarray}
\end{mathletters}
where the decompositions are either given in Table I for the
  ${\mathcal T}$ or in Table II for the ${\mathcal T}^{\mu}$, with
  $\mu\in\{{\rm I,II,III,IV}\}$. The index $l=1,2$ in Eq.~(\ref{o2}) denotes
  the leg on which the observable operates.
\begin{table}[h]
\begin{tabular}{|ccc|}
&$4{\mathcal T}^{\rm I}_{0}$&\\
\hline\hline
$|s\rangle$ & $\longrightarrow$ & $-3|s\rangle$\\
$|t^{i}\rangle$ & $\longrightarrow$ & $|t^{i}\rangle$\\
\hline\hline
&$4{\mathcal T}^{\rm II}_{1}$&\\
\hline\hline
$|s,t^1\rangle$ & $\longrightarrow$ & $|t^1,t^0\rangle-|t^0,t^1\rangle$\\
$|t^1,s\rangle$ & $\longrightarrow$ & $-|t^1,t^0\rangle+|t^0,t^1\rangle$\\
$|s,t^0\rangle$ & $\longrightarrow$ &
         $|t^{1},t^{-1}\rangle+|t^{-1},t^1\rangle$\\ 
$|t^{0},s\rangle$ & $\longrightarrow$ &
         $-|t^1,t^{-1}\rangle+|t^{-1},t^{1}\rangle$\\ 
$|s,t^{-1}\rangle$ & $\longrightarrow$ &
        $|t^0,t^{-1}\rangle-|t^{-1},t^0\rangle$\\ 
$|t^{-1},s\rangle$ & $\longrightarrow$ &
        $-|t^0,t^{-1}\rangle+|t^{-1},t^0\rangle$\\ 
\hline\hline
&${\mathcal T}^{\rm III}_1$&\\
\hline\hline
$|s\rangle$ & $\longrightarrow$ & $|t^0\rangle$\\
$|t^{\pm 1}\rangle$ & $\longrightarrow$ & $0$\\
\hline\hline
&${\mathcal T}^{\rm IV}_0$&\\
\hline\hline
$|s\rangle$ & $\longrightarrow$ & $0$\\
$|t^{i}\rangle$ & $\longrightarrow$ & $i|t^i\rangle$\\
\end{tabular}
\caption{\label{matrix_tab} Action of the local operators 
${\mathcal T}^\mu_i$ appearing in Eqs.~(\ref{ops}). The notation is
the same as in Table I.}
\end{table}

\begin{figure}[h]
  \begin{center}
    \includegraphics[width=6.6cm]{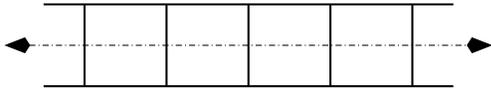}
    \caption{\label{Centerline}The operator ${\mathcal P}$ reflects
      about the depicted axis. A {\it single} rung-singlet (-triplon) has
      odd (even) parity with respect to ${\mathcal P}$. The action of
      ${\mathcal P}$ on the rung-singlet ground state is defined to be
      of even parity ${\mathcal P}|0\rangle=|0\rangle$. If in
      $|0\rangle$ one singlet
      is substituted by a triplon we get the state $|1\rangle$ and
      ${\mathcal P}|1\rangle=-|1\rangle$. Generally, one has ${\mathcal
      P}|n\rangle=(-1)^n|n\rangle$.} 
  \end{center}
\end{figure}

We start with a simple symmetry property. Let ${\mathcal P}$
denote the operator of reflection about 
the center-line of the ladder as depicted in Fig.~\ref{Centerline}.
If $|n\rangle$ denotes a state with $n$
rungs excited to triplons while all other rungs are in the singlet
state we find ${\mathcal P}|n\rangle = (-1)^n|n\rangle$, see caption
of Fig.~\ref{Centerline}. The state $|n\rangle$ might be a linear
combination of many $n$-triplon states so no generality is lost in
writing 
\begin{equation}
  \label{O_d_n}
  {\mathcal O}_{\rm eff}|0\rangle = \sum_{n\ge 0}|n\rangle\ .
\end{equation}
The parity of the ladder observables introduced in Eqs.~(\ref{ops})
with respect to ${\mathcal P}$ is clear from their definition:
${\mathcal O}^{\rm III}$ is odd while ${\mathcal O}^{\rm I}$ and
${\mathcal O}^{\rm IV}$ are even with respect to ${\mathcal P}$, just
as the symmetriezed observable ${\mathcal O}^{\rm II}=({\mathcal
  O}^{\rm II}_{1}+ {\mathcal O}^{\rm II}_{2})/2$. These parities
are conserved under the CUT so that ${\mathcal P}$ applied on both sides
of Eq.~(\ref{O_d_n}) yields 
\begin{equation}
  \label{sym}
  {\mathcal O}_{\rm eff}|0\rangle=\left\{
  \begin{array}{cc}
    \sum_{n}|2n\rangle, & {\mathcal O}_{\rm
    eff}\mbox{ even}\\ 
    \sum_{n}|2n+1\rangle, & {\mathcal O}_{\rm
    eff}\mbox{ odd}
  \end{array}  \right.\ .
\end{equation}
We thus find that an even (odd)
parity of ${\mathcal O}_{\rm eff}$ implies that ${\mathcal O}_{\rm
  eff}$ can inject an even (odd) number of triplons into the system.

The coefficients $c$ in Eq.~(\ref{O_i123_def}) have been calculated
for the one- and two-triplon case on a
computer in a similar fashion as the coefficients $t$ for the
effective Hamiltonian. The implementation of ${\mathcal
  O}_{\rm eff}$ acting on the ground state $|0\rangle$ follows the
same line as described in detail for $H_{\rm eff}$ in
Ref.~\cite{knett99a}. The minimum clusters necessary for some fixed
order arise from the same considerations as in
Sect.~\ref{H_transformation}. Again, the coefficients $c$ are
rational numbers which we computed up to $l_{\rm max}=10{^{\rm th}}$
order for the observables in Eqs.~(\ref{ops}). 

\begin{figure}[h]
  \begin{center}
    \includegraphics[width=8.6cm]{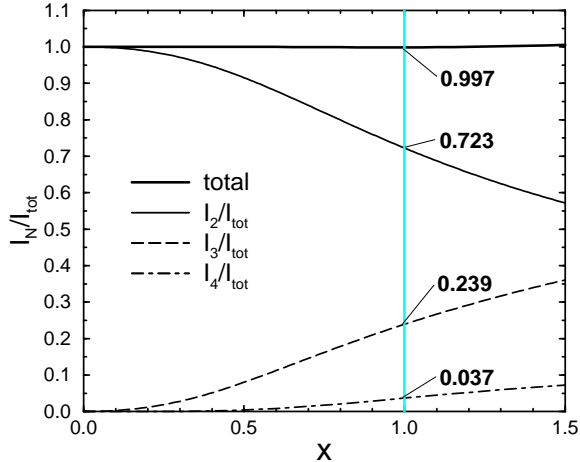}
    \caption{\label{Rel_int}Relative weights for the S=0 operator
      $S_{1,i}S_{1,i+1}$ (Eq.~(\ref{o2})). The $I_{\rm N}$ are calculated
      according to Eq.~(\ref{tot_int}) up to and including order 10, 8
      and 7 in $x$ for the $c_{\{n\}}$ 
      for $N=$2, 3 and 4, respectively. The total
      intensity $I_{\rm tot}$ has been extracted from the 14$^{\rm
      th}$ order result for the ground state energy per spin according
      to Eq.~(\ref{I_tot}). This figure represents an
      improvement of a similar figures in Ref.~\protect\cite{knett01a}. The
      expansion for $I_3$ has been extended by one order and the
      extrapolation of $I_2$ has been improved.}
  \end{center}
\end{figure}

First physically interesting quantities are the {\em spectral weights}
of the observables. As illustrated in Eq.~(\ref{tot_int}) the spectral
weight contained in the $N$-triplon sector, $I_N$, is readily given by
the coefficients $c$. Under certain circumstances the total weight
$I_{\rm tot}=I_0+I_1+I_2+\ldots$ of an observable might be
accessible from sum rules. Let us consider the $S=0$
operator ${\mathcal O}_{\rm eff}^{\rm II}$ in Eq.~(\ref{o2}) as an
example. Here the total weight $I_{\rm tot}(x)$ can be obtained from
the ground state energy per spin 
$\epsilon_0(x)$ given in Eq.~(\ref{epsi_0_plain}). Since $2{\mathcal O}_{\rm
  eff}^{\rm II}(x)= \partial/\partial x H_{\rm eff}(x)$
(cf. Eqs.~(\ref{H_start}) and (\ref{H_pert})) the sum rule can be
  expressed in terms of the effective Hamiltonian, giving rise to  
\begin{align}
  \nonumber 
  I_{\rm tot}&=\sum_{N=0}^{\infty}I_N=\langle 0|{\mathcal O}^2|0\rangle
  - \langle 0|{\mathcal O}|0\rangle ^2\\\label{I_tot}
  &=\frac{3}{16}-\frac{Y}{2}-\frac{Y^2}{2}\ ,
\end{align}
with $Y:=\partial \epsilon_0/\partial x$. If both, $I_{\rm tot}$ and
some of the $I_N$ are known, we can calculate the corresponding {\it
  relative spectral weights} $I_N/I_{\rm tot}$ as functions of
$x$. Fig.~\ref{Rel_int} shows the resulting relative weights for the
observable ${\mathcal O}^{\rm II}_{\rm eff}$ for the first four triplon 
sectors. 
Since one cannot form an $S=0$ object from a single rung-triplon there
is no $I_1$ for this observable. The contribution of $I_5/I_{\rm tot}$
is of order 10$^{-3}$ leading to no visible changes in
Fig.~\ref{Rel_int}. Contributions of higher triplon channels are
expected to be even smaller.    
All relative weights add up to unity. As
can be seen in Fig.~\ref{Rel_int} the first four relative weights
fulfill this requirement with great precision. For $x=0$ the singlet
made from two isolated-rung triplons contains the full weight of the
considered operator. As $x$ increases the triplons start to polarize
their environment, the two-triplon weight decreases and
multi-triplon states gain weight. A similar figure for the
$S=1$ operator $S_{1,i}^z$ can be found in Ref.~\cite{knett01a}
(Fig.~2). 

From the depicted result we conclude that the triplon is an excellent
choice for quasi-particle in the ladder system. For $x$ not
too large most of the spectral weight is captured by a few
triplons. Calculations containing only a few triplons suffice to
explain most of the physics for $x\lesssim 1.5$.

\begin{figure}[h]
  \begin{center}
    \includegraphics[width=8.6cm]{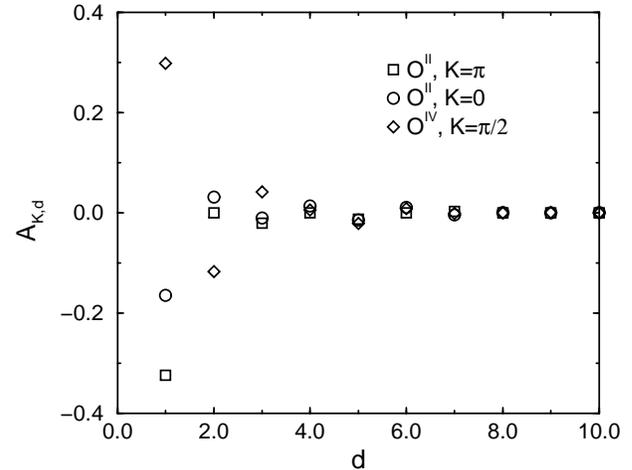}
    \caption{\label{Ak_d-dropoff} The two-triplon momentum dependent
      coefficients $A_{K,d}$
      of the observables ${\mathcal O}^{II}$ and ${\mathcal O}^{IV}$
      for all calculated distances $d$ at $x=1$ and momenta as
      indicated. The coefficients rapidly drop to zero with increasing
      distance. It is not necessary to go to larger orders,
      i.e., distances.}    
  \end{center}
\end{figure}

Eqs.~(\ref{O_K_3}) and (\ref{O_K_4}) show how the momentum depend
coefficients $A_k$ (one-triplon) and $A_{K,d}$ (two-triplons) can be
calculated from the corresponding $c$-coefficients. 
For the two-triplon sector we provide some examples to 3$^{\rm rd}$ order in 
$x$. The shown coefficients $A_{K,d}$ belong to the symmetrized observable
${\mathcal O}^{\rm II}=({\mathcal O}^{\rm II}_1+{\mathcal O}^{\rm
  II}_2)/2$ or to the observable ${\mathcal O}^{\rm IV}$, respectively,
\begin{align}
  \nonumber
  A_{K,1}^{\rm II}&=-\frac{1}{4}-\frac{1}{8}\,x+\frac{1}{8}\,
    \left(\frac{5}{8}\,\cos \left( K \right) 
    +\frac{5}{8} \right) {x}^{2}\\\nonumber
  &+\frac{1}{8}\, \left( {\frac {25}{16}}+{\frac {17}{16}}\,\cos \left( K
    \right)  \right) {x}^{3}\\\nonumber
  A_{K,2}^{\rm II}&=\frac{1}{8}\,\cos \left( \frac{1}{2}\,K \right)
  x+\frac{1}{16}\,\cos \left(
    \frac{1}{2}\,K \right) {x}^{2}\\\nonumber
  &+\frac{1}{16}\, \left( -{\frac
      {37}{16}}\,\cos \left( \frac{1}{2}\,K \right) -{\frac
      {13}{16}}\,\cos \left( \frac{3}{2}\,K \right)  \right) {x}^{3}\\
  A_{K,1}^{\rm IV}&=\frac{1}{2}\,x\sin \left( \frac{1}{2}\,K \right)
  +\frac{1}{4}\,{x}^{2}\sin \left( \frac{1}{2}\,K
  \right)\\
  &-{\frac {11}{64}}\,{x}^{3}\sin \left( \frac{1}{2}\,K \right)\ .
\end{align}
Fig.~\ref{Ak_d-dropoff} displays the coefficients of ${\mathcal O}^{II}$
and ${\mathcal O}^{IV}$ as function of the relative distance $d$ for
fixed $K$-values at $x=1$. The depicted values are obtained by using
standard Pad\'e techniques for the $A_{K,d}$ as polynomials in $x$ for
fixed momentum $K$. 
All calculated one-triplon ($A_k$) and two-triplon
($A_{K,d}$) coefficients will be made available on our home
pages~\cite{homepages}.

\section{Evaluating the Green Function}
\label{sec_G_function}
We are now in the position to calculate the zero temperature one- and
two-triplon 
spectral densities associated with the ladder observables introduced
in the last section. To this end we start by analyzing the energy and
momentum resolved retarded
zero temperature Green function
\begin{align}
  \nonumber
  {\mathcal G}^{\mathcal O}(K,\omega)&=\\\label{GREEN_DIE_ERSTE}
  &\hspace*{-15mm}\left\langle \psi_0\left|{\mathcal
  O}^{\dagger} (K)\frac{1}{\omega-\left(H(K)-E_0\right)+i0^+}{\mathcal
  O}(K)\right|\psi_0 \right\rangle\ ,
\end{align}
from which the spectral density ${\mathcal S}(K,w)$ follows by taking
the negative imaginary part. As explained in Sect.~I.B, all operators
can be replaced by their effective counterparts after the transformation
and the ground state $|\psi_0\rangle$ by the triplon vacuum $|0\rangle$.

\subsection{One-Triplon Green Function}
In the one-triplon case the calculation is particularly simple. Using
Dirac's identity 
\begin{equation}
  \frac{1}{x-x_0\pm i0^+}={\mathcal P}\frac{1}{x-x_0}\mp
  i\pi\delta(x-x_0)\ ,
\end{equation}
where ${\mathcal P}$ denotes Cauchy's principal value, we find 
\begin{align}
  \nonumber {\mathcal S}(k,\omega)&=\left\langle 0 \left|{\mathcal
  O}^{\dagger}_{1,0}(k) \delta(\omega-H_1){\mathcal
  O}_{1,0}^{\phantom\dagger}(k)\right|0\right\rangle\\\nonumber &=
  |A_{k}|^2\left\langle k\left|\delta(w-H_1)\right|k\right\rangle\\
  &= |A_{k}|^2\delta\left(\omega-\omega(k)\right)\ .
\end{align}
The one-triplon dispersion $\omega(k)$ and the observable
coefficient $A_{k}$ are readily given by Eqs.~(\ref{1-energ}) and
(\ref{O_K_3}), respectively. At each point $(k,\omega(k))$ the 
corresponding weight is given by the square of the modulus of
$A_{k}$ which 
is a polynomial in $x$. The result is thus obtained by
assigning a $\delta$-function with corresponding weight to each point
$(k,\omega(k))$.

\subsection{Two-Triplon Green Function}
For the two-triplon case we choose to evaluate the effective Green
function by tridiagonalization. This leads to the continued fraction
expression (\cite{zwanz61,mori65b,gagli87}, for overviews see
Refs.~\cite{petti85,viswa94})
\begin{equation}
  \label{confrac_THE_FIRST}
  {\mathcal G}_{2,0}^{\mathcal O} (K,\omega)=
  \frac{\langle 0| {\mathcal O}^{\dagger}_{2,0}(K){\mathcal
  O}_{2,0}^{\phantom\dagger}(K)|0\rangle}{\omega-a_0-{\displaystyle
  \frac{b_1^2}{\omega-a_1-{\displaystyle \frac{b_2^2}{\omega
  -\cdots}}}}}\ ,
\end{equation}
where we can also write $\sum_{d}|A_{K,d}|^2$ for the
expression in the numerator on the right hand side.
The coefficients $A_{K,d}$ are given by Eq.~(\ref{O_K_4}).
The coefficients $a_i$ and $b_i^2$ are calculated by repeated application
of $H_{\rm eff}-H_0=H_1+H_2$ on the initial two-triplon momentum
state $|{\rm Init}\rangle=|f_0\rangle={\mathcal
  O}_{2,0}(K)|0\rangle$. The action of $H_1$ and $H_2$ on these states
have been calculated previously. The results are given in Eqs.~(\ref{T}) and
(\ref{W_2}), respectively.

Setting the state
$|f_{-1}\rangle=0$ the recursion (Lanczos tridiagonalization) 
\begin{equation}
  |f_{n+1}\rangle = (H_1+H_2)|f_n\rangle -a_n|f_n\rangle
  -b_n^2|f_{n-1}\rangle\ \ ,\ n\in{\mathbb N}
\end{equation}
generates a set of orthogonal states $|f_n\rangle$ if the
coefficients are defined according to
\begin{equation}
  \nonumber a_n = \frac{\langle f_n|(H_1+H_2)|f_n\rangle}{\langle
  f_n|f_n\rangle}\ ,\ \ \ \ \ b_{n+1}^2 = \frac{\langle
  f_{n+1}|f_{n+1}\rangle}{\langle f_n|f_n\rangle}\ .
\end{equation}
In the generated $\{|f_n\rangle\}$-basis $H_{\rm eff}$ is a
  tridiagonal matrix, where the $a_i$ are the diagonal matrix elements
  and the $b_i$ are the elements on the second diagonal. All other
  matrix elements are zero.

Fig.~\ref{T_W_SV_matrix} illustrates the   
procedure for the two-triplon sector. For fixed $K$ the relative
(positive) distance $d$ between the two injected triplons is the
only remaining quantum number. In this basis $H_1+H_2$ is
represented as a matrix (left side). The matrix elements are
polynomials in the perturbation parameter $x$. We have to apply this
matrix iteratively to the $|f_n\rangle$. The components $A_{K,d}$ of
the initial vector $|f_0\rangle=|{\rm Init}\rangle$ are polynomials in
$x$ for fixed $K$. By this procedure a new basis $|f_n\rangle$ is
generated in which the fairly complicated matrix in
Fig.~\ref{T_W_SV_matrix} is simplified to a tridiagonal one.

 The general case of more
than two triplons can be treated similarly. For
$n$ triplons we have to consider the conserved total
momentum $K$ and 
$n-1$ relative distances. Then, for fixed $K$, $|{\rm Init}\rangle$
and $H_{\rm eff}$ are still represented by a vector and matrices,
respectively. But they are more complicated. For three
triplons we obviously 
have to apply $H_1+H_2+H_3$ to $|{\rm Init}\rangle$. For four
triplons $H_4$ is added and so on. The calculation of $H_n$ with
$n>2$ is indicated in Ref.~\cite{knett03a}. 

Thus, the full many-particle problem is effectively reduced
to a few-particle problem! Further, after fixing the parameter value
$x$ the coefficients $a_i$ and $b_i$ are obtained by the numerical
Lanczos tridiagonalization, which is left to the computer.
This procedure is realized for fixed $x$ and $K$. For the
spin ladder we were able to implement a maximum relative distance $d$ of
$\approx 10000$, allowing to repeat the recursion about 650 times
giving the first 650 coefficients $a_i$ and $b^2_i$.

Thus the chosen method to evaluate
the effective Green function introduces no quantitative finite size
effects. The problem of calculating the spectral densities for given
effective Hamiltonians and observables comprises the two quantum numbers
$K$ and $d$ in the two-triplon sector. For each triplon
more, there is one more relative distance to be considered, see
above. Our calculations are exact and without finite-size effect to
the given order for the total momentum $K$.

There are two approximations that involve the relative distance $d$.
They will be discussed in the
following. The prevailing approximation is caused by the finiteness
of the perturbative calculations. The true many-triplon interactions are
accounted for only if all involved particles are within a certain
finite distance to each other. This approximation is controlled, since
one generically observes a rather sharp drop of the interaction
matrix elements with increasing distances (see Fig.~\ref{H_2_dropoff} for the
spin ladder). Gapped systems with
finite correlation lengths are well suited to be
tackled by our method. Difficulties arise if the 
correlations drop 
slowly with increasing distances. In this case the truncations in
real-space might not be justified.

A second minor approximation to the results involving the
relative distance $d$ is introduced by truncating the continued 
fraction expansion of the Green function.  However, allowing for distances
of up to 10000 lattice spacings as in the ladder example guarantees that this
error is extremely small in comparison to the error introduced by 
truncating the perturbative expansion as discussed in the preceding paragraph.

The finiteness of the continued fraction can
be partly compensated by suitable terminations as shall be explained in
the following subsection.

\subsection{Terminator}
The spectral density ${\mathcal S}(K,\omega)$ at fixed $K$ as obtained
from the truncated continued fraction~(\ref{confrac_THE_FIRST})
of the effective Green function has poles at the zeros
of the denominator. Thus, ${\mathcal S}$ would be a collection
of sharp peaks. A slight broadening of ${\mathcal S}$ via
$\omega\rightarrow\omega+i\delta$ ($\delta$ small) in ${\mathcal G}$
will smear out all poles to give a continuous function for all
practical purposes. However, we can achieve perfect resolution of
${\mathcal S}$ as continuous function by introducing a
proper termination of the continued fraction exploiting the
one-dimensionality of the considered model.

For fixed total momentum $K$ the (upper)
lower band edges ($\epsilon_{\rm ub}$) $\epsilon_{\rm lb}$ of the
two-triplon continuum can be calculated from the one-triplon
dispersion $\omega_1$~(\ref{1-energ}). All energies of the two-triplon
continuum are seized by
\begin{equation}
  \label{omega_II}
  \omega_2(K,q)=\omega_1\left(K/2+q\right)
  +\omega_1\left(K/2-q\right)\ ,
\end{equation}
where $q\in [-\pi,\pi]$ denotes the relative momentum. Therefore, we
can calculate $\epsilon_{\rm ub}$ and $\epsilon_{\rm lb}$ from the
one-triplon dispersion
\begin{align}
  \nonumber \epsilon_{\rm ub}(K)&=\max_{q}(\omega_2(K,q))\\\label{ucl}
  \epsilon_{\rm lb}(K)&=\min_{q}(\omega_2(K,q))\ .
\end{align}
For fixed $K$ the upper and lower band edges $\epsilon_{\rm ub}$
and $\epsilon_{\rm lb}$ determine the values to which the continued
fraction 
coefficients $a_i$ and $b_i$ converge for
$i\rightarrow\infty$. One finds $a_{\infty}= (\epsilon_{\rm
  ub}+\epsilon_{\rm lb})/2$ and $b_{\infty}= (\epsilon_{\rm
  ub}-\epsilon_{\rm lb})/4$ \cite{petti85}. This serves as an
independent check for the calculated coefficients. 

If we assume the system under study to be gapped 
the massive elementary excitations show quadratic behaviour at the
dispersion extrema. Then it is generic that $\omega_1(q)$ is smooth,
i.e., two-fold continuously differentiable, and so is $\omega_2(K,q)$.
Since the problem is one-dimensional
 there are square-root singularities in the density 
of states at the edges of the continuum if the two particles are  
asymptotically free at large distances. 
In conclusion, a square root termination for the
continued fraction is appropriate~\cite{petti85,viswa94}. The listed
properties lead to a convergent behaviour of $a_i$ and $b_i^2$ with 
\begin{eqnarray}
  \nonumber
  a_i&=&a_{\infty}+{\mathcal O}(1/i^3)\\
  b_i&=&b_{\infty}+{\mathcal O}(1/i^3)\ ,
\end{eqnarray}
and it is well justified to assume $a_i$ and $b_i$ to be constant
beyond a certain (large) fraction depth $i$. 
Hence we use
\begin{equation}
  \label{D-def}
  D=4b^2_{\infty}-(\omega-a_{\infty})^2
\end{equation}
and define
\begin{align}
  \nonumber {\tau}&=\frac{1}{2b^2_{\infty}}
  \left(\omega-a_{\infty}-\sqrt{-D}\right)\ \mbox{for}\ 
  \omega\ge \epsilon_{\rm ub}
  \\\nonumber {\tau}&=\frac{1}{2b^2_{\infty}}
  \left(\omega-a_{\infty}-i\sqrt{D}\right)\ \mbox{for}\ 
  \epsilon_{\rm lb} <\omega < \epsilon_{\rm ub}
  \\\label{term} {\tau}&=\frac{1}{2b^2_{\infty}}
  \left(\omega-a_{\infty}+\sqrt{-D}\right)\ \mbox{for}\ 
  \omega \le \epsilon_{\rm lb}
\end{align}
The last calculated $b_i^2$ in Eq.~(\ref{confrac_THE_FIRST}) is
multiplied by the appropriate terminator $\tau$.
Taking the imaginary part of the resulting
expression for the case within the continuum yields the continuous
part of the spectral density ${\mathcal S}$ in the thermodynamic
limit. The result is a continuous function
displaying the full weight of the continuum correctly, limited only by
the finite order of the series expansion.

In the case of bound states the Green function can be
written as ($K$ is assumed fixed)
\begin{equation}
  \label{G_bound}
  {\mathcal G}^{\mathcal O}(\omega)=\frac{\langle 0|
  {\mathcal O}_{2,0}^{\dagger}{\mathcal
  O}_{2,0}^{\phantom\dagger}|0\rangle}{\omega-f(\omega)}\ , 
\end{equation}
where the function $f(\omega)$ is real-valued for
$\omega \le \epsilon_{\rm lb}$. The position of possible bound
states is given by the zeros of $g(\omega)=\omega-f(\omega)$. Let
$\omega_0$ be such a zero of $g$. We expand $g$ about $\omega_0$ in
$\omega-\omega_0$ to first order which is sufficient for small
deviations from $\omega_0$
\begin{equation}
  \label{G_bound_II}
  {\mathcal G}^{\mathcal O}(\omega)\approx\frac{\langle 0|
  {\mathcal O}_{2,0}^{\dagger}{\mathcal
  O}_{2,0}^{\phantom\dagger}|0\rangle}{(\omega-\omega_0) 
  (1-{\partial_{\omega}}f(\omega_0))}\ .
\end{equation}
If ${\mathcal G}^{\mathcal O}$ is the retarded Green
function the Dirac-identity yields
\begin{equation}
  \label{S_bound}
  {\mathcal S}(\omega)|_{\omega\approx\omega_0}=-\frac{1}{\pi}{\rm Im}{\mathcal G}^{\mathcal
  O}(\omega)=\frac{\langle 0| {\mathcal
  O}_{2,0}^{\dagger}{\mathcal O}_{2,0}^{\phantom\dagger}|0\rangle} 
  {1-{\partial_{\omega}}f(\omega_0)}\delta(\omega-\omega_0)\ ,
\end{equation}
clarifying that a possible bound state shows up as a
$\delta$-function. Its spectral weight is given by
\begin{equation}
  \label{pos_weight}
  I_{\rm bound}^{-1}={\partial_{\omega}}\left({\mathcal G}^{\mathcal
    O}(\omega)^{-1}\right)|_{\omega=\omega_0}= 
   \frac{1-{\partial_{\omega}}f(\omega_0)}{\langle 0| {\mathcal
  O}_{2,0}^{\dagger}{\mathcal O}_{2,0}^{\phantom\dagger}|0\rangle} 
  \ ,
\end{equation}
which is easy to calculate once the continued fraction is known.

The methods explained in this section have been used to derive the
spectral densities presented in earlier publications; see
Refs.~\cite{gruni02c,knett01a,knett02b,schmi01a,schmi03b,schmi03c,windt01,knett03b}.
Finally we address the necessary extrapolations if the perturbation
parameter $x$ is not small.

\section{Optimized perturbation theory}
\label{OPT_section}
The results for the one-triplon dispersion $\omega(k,x)$ and the
matrix elements of $H_1$ and $H_2$ in the two-triplon sector are
perturbative. They rely on 
effective operators  calculated as truncated series, i.e., polynomials,
in $x$. The theory is controlled in the sense that it is correct for
$x\rightarrow 0$. But in general we do not have information about the radius
of convergence. A standard way to extrapolate the polynomials is the
use of approximants like Pad\'e - or Dlog-Pad\'e - approximants and
others~\cite{domb83_pade}. This is a feasible task if one is dealing
with a {\it few} quantities only. However, for the matrices $H_1$ and
$H_2$ of Sect.~II there are more than 100 matrix elements, each a
truncated series in $x$, to be extrapolated for each $K$. Clearly,
this task has to be automatized. The Pad\'e-methods do not allow a
simple automatization, since the resulting approximants are not
sufficiently robust. Some of the possible Pad\'e approximants to a
given polynomial might display spurious singularities and there is no
way to predict this to happen. Pad\'e approximants need to be inspected
manually. 

Some progress can be made by a recently developed extrapolation
procedure introduced in Ref.~\cite{schmi02}. This technique relies on
re-expanding the 
initially obtain truncated series expansion in the external
perturbation parameter $x$ by a suitable internal parameter $p(x)$. The
latter is chosen so that it comprises information on special
system-dependent behaviour (such as tendencies in the vicinity of
system-specific singularities) to which the external parameter is not
sensitive. This method has been used successfully in
Ref.~\cite{schmi03e} to calculate the transition line between the
rung-singlet phase and a spontaneously dimerized phase for the spin
ladder system including ring exchange. 

In this article we propose optimized
perturbation theory (OPT), which is based on the principle of minimal
sensitivity~\cite{steve81}, as a particularly robust technique to
simultaneously extrapolate a large number of polynomials in an
automatized fashion.

For the general derivation of OPT we go back to the 
beginning of our perturbational approach where we assumed that the
Hamiltonian can be split into an unperturbed part $U$ and a
perturbation $V$
\begin{equation}
\label{OPT_begin_1}
  H(x)=U+xV\ .
\end{equation}
The fundamental idea of optimized perturbation theory (OPT) is to 
modify this splitting and to adjust this modification by an additional
control parameter $a$
\begin{align}
  \label{OPT_begin_2}
  H(x;a)&=(1+a)U+xV - aU\\\nonumber &=
  (1+a)\tilde{H}(\tilde{x};\tilde{a})\\\nonumber 
  \tilde{H}(\tilde{x};\tilde{a})&=U+\tilde{x}(V+\tilde{a}U),\ \tilde{x}=\frac{x}{1+a},\ \tilde{a}=-\frac{a}{x}\
  .
\end{align}
We consider $\tilde x$ to be the new expansion parameter.  The
Hamiltonian $H(x;a)=(1+a)\tilde{H}(\tilde{x};\tilde{a})$ is identical
to Hamiltonian $H(x)$. Hence no energy depends on $a$ in the exact
result. Let us consider the gap $\Delta(x)$ as generic example. It
does not depend on $a$. But the truncated series expansion $\Delta_{\rm 
trunc}(x;a)$ resulting from $H(x;a)$ will depend on $a$. 
We at least demand stationarity in this
parameter. This is motivated by the idea of minimal sensitivity
\cite{steve81}. We write
\begin{equation}
  \label{DELTA_TRUNC_1}
 \Delta_{\rm trunc}(x;a)=(1+a){\mathcal
T}\!\!\overset{n}{\underset{\tilde{x}=0}{\mbox{\LARGE $|$}}}
\tilde{\Delta}(\tilde{x};\tilde{a})\ ,
\end{equation}
where $\tilde{\Delta}(\tilde{x};\tilde{a})$ denotes the energy
resulting from
$\tilde{H}(\tilde{x};\tilde{a})$ and ${\mathcal
T}\!\!\overset{n}{\underset{x=x_0}{\mbox{\large $|$}}}f(x)$ is the
$n^{\rm th}$ order Taylor expansion of $f(x)$ in $x$ about $x=x_0$.
 Requiring  stationarity  leads to the criterion
\begin{equation}
  \label{OPT_criterion}
  \partial_{a}\Delta_{\rm trunc}(x;a)|_{a=a_{\rm opt}}=0\ .
\end{equation}
In general, $\Delta_{\rm trunc}(x;a_{\rm opt})$ converges faster than
the corresponding series expansions of $\Delta(x)$, since the
additional degree of freedom can be used to optimize the splitting into
an unperturbed and a perturbing part~\cite{steve81}. In other words, 
the system has  the freedom to choose the best splitting depending on 
the series under study. Moreover, in some cases a convergent
series expansion can be enforced by OPT even 
if the original series diverges. In Ref.~\cite{steve81}, see also
references therein, the harmonic oscillator perturbed by a quartic 
potential is given as an example whose standard series expansion for
the ground state energy diverges~\cite{bende69}.

To be more specific, the series of
$\tilde{\Delta}(\tilde{x};\tilde{a})$ is needed. We rewrite 
\begin{align}
  \nonumber
  \tilde{H}(\tilde{x};\tilde{a})&=U+\tilde{x}(V+ \tilde{a}U)\\\label{H_zwi_2}
  &=(1+\tilde{a}\tilde{x}) \left[ U+
  \frac{\tilde{x}}{1+\tilde{a}\tilde{x}} V\right]\ .
\end{align}
One clearly sees that the series of $\tilde{\Delta}$ in $\tilde{x}$
can be obtained by expanding the expression 
\begin{equation}
  \tilde{\Delta}(\tilde{x};\tilde{a})=(1+\tilde{a}\tilde{x})
  \Delta\left(\frac{\tilde{x}}{1+\tilde{a}\tilde{x}}\right)\ ,
\end{equation}
in $\tilde{x}$.
Let us assume that we had already calculated $\Delta(x)$ as a truncated
series $\Delta_{\rm trunc}(x)$ from $H_{\rm eff}$. Then $\Delta_{\rm
  trunc}(x;a)$ is obtained by
\begin{align}
   \Delta_{\rm trunc}(x;a)&=(1+a) {\mathcal
  T}\!\!\overset{n_{\rm max}}{\underset{\tilde{x}=0}{\mbox{\huge
  $|$}}} \tilde{\Delta}(\tilde{x};\tilde{a}) \\\nonumber &=(1+a){\mathcal
  T}\!\!\overset{n_{\rm max}}{\underset{\tilde{x}=0}{\mbox{\huge
  $|$}}}\!\!\!\left\{ (1+\tilde{a}\tilde{x})\Delta_{\rm
  trunc}\left(\frac{\tilde{x}}{1+\tilde{a}\tilde{x}}\right)\right\}\ ,
\end{align}
where $n_{\rm max}$ is the maximum order to which we obtained
$\Delta_{\rm trunc}(x)$ before. Finally $\tilde{x}$ and $\tilde{a}$
are re-substitute  by their definitions in Eq~(\ref{OPT_begin_2}). 

In order to do all steps in one we
introduce an auxiliary variable $\lambda$ for the derivation. Then, the
Taylor expansion in $\tilde{x}$ can be replaced by an expansion in
$\lambda$
\begin{equation}
  \tilde{x}=\frac{\lambda x}{1+a}\ \mbox{ and }
  \tilde{a}=-\frac{a}{x}\ ,
\end{equation}
where it is understood, that the final result is obtained for
$\lambda=1$. This leads to
\begin{align}
  \label{DELTA_OPT_FIN}
  \Delta_{\rm trunc}(x;a)&=\\\nonumber
   & \hspace*{-15mm}\left[{\mathcal T}\!\!\overset{n_{\rm
  max}}{\underset{\lambda=0}{\mbox{\huge $|$}}}
  (1+a(1-\lambda))\Delta_{\rm trunc}\left(\frac{\lambda
  x}{1+a(1-\lambda)}\right)\right]_{\lambda=1}\ .
\end{align}

The bottomline is that OPT can be used without computing any new
coefficients. Only some straightforward computer algebra is needed. 
We take the direct expansions for the energy
$\Delta_{\rm trunc}(x)$ obtained from the effective Hamiltonian and
substitute and re-expand according to Eq.~(\ref{DELTA_OPT_FIN}) to get the
optimized expansions $\Delta_{\rm trunc}(x;a_{\rm opt})$ where
$a_{\rm opt}$ is given by the minimal sensitivity
criterion~(\ref{OPT_criterion}).

From the discussion above it is clear that also other quantities
$A_{\rm trunc}(x)$ like matrix elements of effective observables
 can be optimized analogously
\begin{align}
  \label{REST_OPT_FIN}
  A_{\rm trunc}(x;a)&=\\\nonumber
  & \hspace*{-15mm}\left[{\mathcal T}\!\!\overset{n_{\rm
  max}}{\underset{\lambda=0}{\mbox{\huge $|$}}} A_{\rm
  trunc}\left(x\rightarrow\frac{\lambda
  x}{1+a(1-\lambda)}\right)\right]_{\lambda=1}\ .
\end{align}
The prefactor $(1+a(1-\lambda))$ is dropped since $A$ does not depend
on the global energy scale in $H$ in contrast to $\Delta$. Note that
formula (\ref{DELTA_OPT_FIN}) can be used for any matrix element of
the Hamiltonian. Formula~(\ref{REST_OPT_FIN}) can be used for any
dimensionless matrix element of an observable. 

The criterion of minimal sensitivity allows to  elaborate further on
the structure of $a_{\rm opt}$. We will show that 
\begin{equation}
  \label{assert_opt}
  a_{\rm opt}=\alpha_{\rm opt}x\ ,
\end{equation}
holds.
Let $R_{\rm trunc}(x;a)$ be the truncated series expansion of the
quantity for which we want to find the optimum value $a_{\rm
opt}$. In the following discussion $R$ is an energy $\Delta$ or
some matrix element $A$. To ease the notation we introduce the
function 
\begin{equation}
  g(u,v)=
  \begin{cases}
    v\Delta_{\rm trunc}(u/v) & \text{for energies},\\
    A_{\rm trunc}(u/v)  & \text{for matrix elements.}
  \end{cases}
\end{equation}
The derivative of $g$ with respect to $v$ is denoted by
$f(u,v)=\partial_v g(u,v)$. The problem of calculating $a_{\rm
  opt}$ reduces to 
\begin{align}
\nonumber 0 \dot{=} \partial_a R_{\rm trunc}(x;a) &= {\mathcal
  T}\!\!\overset{n}{\underset{\lambda=0}{\mbox{\huge $|$}}} \partial_a
  g(\lambda x, 1+a(1-\lambda))\\ &= {\mathcal
  T}\!\!\overset{n}{\underset{\lambda=0}{\mbox{\huge $|$}}} f(\lambda
  x, 1+a(1-\lambda))(1-\lambda)\ ,
\end{align}
where the notation for $\lambda=1$ in the end is suppressed for clarity.
For the following argument it is important to see that
\begin{align}
  \label{F_TAYLOR_EXPAND}
  &{\mathcal T}\!\!\overset{n}{\underset{\lambda=0}{\mbox{\huge $|$}}}
  f(\lambda x, 1+a(1-\lambda))(1-\lambda) =\\\nonumber
  &f_n\lambda^n+(1-\lambda)
  {\mathcal T}\!\!\overset{n-1}{\underset{\lambda=0}{\mbox{\huge
  $|$}}} f(\lambda x, 1+a(1-\lambda))\ ,
\end{align}
holds, 
where $f_n$ denotes the $n^{\rm th}$ coefficient in the Taylor
expansion of $f$ with respect to $\lambda$
\begin{equation}
\label{fn_structure}
  f_n=\frac{1}{n!}(\partial_{\lambda})^n f(\lambda x, 1+a(1-\lambda))\
  .
\end{equation}
For the final value $\lambda=1$, the second term on the right hand side
of Eq.~(\ref{F_TAYLOR_EXPAND}) vanishes. In addition, the structure in
Eq.~(\ref{fn_structure}) is
such, that with each derivation with respect to $\lambda$ we obtain
either an $x$ or an $a$ as internal derivative of the chain rule. Thus,
setting $\lambda=1$ in 
the end, we find 
\begin{equation}
\partial_aR_{\rm trunc}(x;a)=f_n
\end{equation}
to be a {\it homogeneous}
polynomial in the variables $x$ and $a$. In an $n^{\rm th}$ order
expansion the criterion of minimal sensitivity reads
\begin{equation}
 0 \dot{=} \partial_a R_{\rm trunc}(x;a)|_{a=a_{\rm opt}}=
\sum_{i=0}^n R_ia^ix^{n-i}|_{a=a_{\rm opt}}\ ,
\end{equation}
which clearly shows, that we can always write $a_{\rm opt}= \alpha_{\rm
  opt}x$. This proves the assertion~(\ref{assert_opt}).

The proposed OPT procedure can be performed for all physical
quantities of interest in particular for the matrix elements of $H_1$
and $H_2$.  No new calculations are required. Instead, the plain
series results can be promoted to OPT results by simple
substitutions and re-expansion.

\begin{figure}[h]
  \begin{center}
    \includegraphics[width=8.6cm]{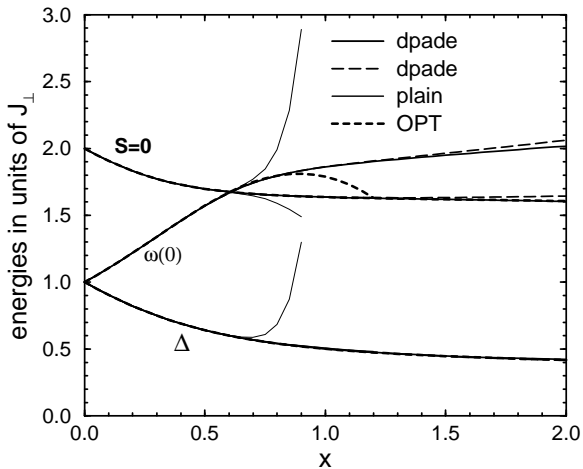}
    \caption{\label{OPT_Vergl} The elementary triplon gap
      $\Delta=\omega(k=\pi)$, $\omega(k=0)$ and the $S=0$ two-triplon gap
      as functions of the perturbation parameter $x$. For all energy
      levels the validity of the plain series results (thin lines)
      starts to break down at $x\approx 0.6$. Various biased
      Dlog-Pad\'e approximants (dpade in legend) are
      shown for each quantity as thick 
      solid or long dashed lines. They yield the most reliable
      extrapolations. The results obtained from optimized perturbation
      theory (OPT) with $\alpha_{\rm opt}=2.9x$ are depicted as thick short
      dashed lines. Except for $\omega(k=0)$ there is no visible
      deviation from the Dlog-Pad\'e results. For $x \le 1$ the
      optimized results can be used without appreciable loss of accuracy.}   
  \end{center}
\end{figure}

In the application we modify the OPT idea slightly. We assume that
there is an optimum splitting for each order, that means an optimum value for
$\alpha_{\rm opt}$, to do the perturbation expansion. This means that
we do {\it not} adapt $\alpha_{\rm opt}$ to various different
quantities, but we use one universal value depending only on the order of
the series. This  value is
determined by simultaneously optimizing some simple quantities like the
one-triplon gap $\Delta(x) =\omega(x,k=\pi)$ (Eq.~(\ref{1-energ})),
the $S=0$ bound state energy at $K=\pi$ ($\Delta_{S=0}(x)$) and
others with respect to the best Dlog-Pad\'e approximant of these
quantities. This approach is based on the plausible assumption that all
considered quantities (here: energy levels) are governed by the same
singularities reflecting the underlying physics. This idea is supported by
the fact that all 
energy expansions we obtained start to deviate from their best
extrapolations at about the same value for $x$
(cf.~Fig.~\ref{OPT_Vergl}). Thus, in contrast to the original spirit of the OPT
method, we propose that $\alpha_{\rm opt}$ essentially depends on the
model and the order of the expansions only, but not on the particular
quantity under study.

In Fig.~\ref{OPT_Vergl}, $\Delta(x)=\omega(x,k=\pi)$,
$\omega(x,k=0)$ and $\Delta_{S=0}(x)$ are plotted as
functions of $x$.
The thin solid lines correspond to the plain
expansion results, reliable up to $x\approx 0.6$. Various 
Dlog-Pad\'e approximants for each energy are depicted by thick solid
and thick long-dashed lines. They are biased by including the fact,
that all energies in the ladder system should grow linearly in $x$ for
$x\rightarrow \infty$. In that limit the system maps onto 
two decoupled $S=1/2$ chains whose energies $E_{\nu}$, measured in
units of the remaining coupling constant, are
constants $E_{\nu}/J_{\parallel}$=const..
Measuring these energies in units of $J_{\perp}$, as we do, stipulates
$E_{\nu}/J_{\perp}=E_{\nu}x/J_{\parallel}\Rightarrow
E_{\nu}/J_{\perp}\sim x$ for $x\gg 1$ giving rise to the
extrapolation bias. The thick short-dashed
lines show the corresponding OPT-results with $\alpha_{\rm
  opt}=2.9x$. The figure illustrates that it is possible to choose a
fixed  $\alpha_{\rm opt}$ leading to a global improvement in all
energy levels. For some levels the improvement is very good
(e.g. $\Delta_{S=0}(x)$ or $\Delta(x)$), for others it is still
reasonable good (e.g. $\omega(x,0)$).

The value for $\alpha_{\rm opt}$ depends on the order of the original
truncated series. We found that $\alpha_{\rm
  opt}=2.6x$ gives best results for 13$^{\rm th}$ order
expansions. Thus, matrix elements of $H_1$ are optimized with 
$\alpha_{\rm opt}=2.9x$ and those of $H_2$ with $\alpha_{\rm opt}=2.6x$.

It is a significant advantage that the OPT procedure as we use it
is linear. Let $O_{\alpha}[\cdot]$ denote the OPT procedure such that
\begin{equation}
  f(x;a_{\rm opt})=O_{\alpha} [f(x)]
\end{equation}
is the optimized series obtained from the direct series $f(x)$. Then,
for a linear quantity $F(x)=\sum_i a_i f_i(x)$ one has
\begin{align}
  \nonumber F(x;a_{\rm opt}) &= O_{\alpha}\left[\sum_i a_i f_i(x)\right]\\ &=
  \sum_i a_i \ O_{\alpha}\left[f_i(x)\right]\ ,
\end{align}
as long as all $f_i$ are given to the same order.
For the two-particle interaction part $H_2$ of $H_{\rm eff}$, for
instance, this means that one can 
choose to optimize the matrix elements $\langle K,d'|H_2|K,d\rangle$
directly, or to 
optimize the two-particle interaction coefficients $t_{d;r,d'}$ before the
sums of the Fourier transform is carried out (see Eq.~\ref{W_2}). This
linearity is not ensured by Pad\'e or Dlog-Pad\'e extrapolations which
represents a serious caveat in the practical use. 

We like to stress that OPT does not yield the best approximants
one can think of. It is rather a
compromise between feasibility and quality. The OPT
method represents a very robust and smooth approximation scheme in the
sense that none of the approximants diverges or produces unexpected
pathologies. Its linearity makes it particular appropriate for the
treatment of Fourier transformed matrix elements. 

Some of the matrix elements of $H_1$ and $H_2$
have been cross-checked with (Dlog-)Pad\'e approximants leading to the
conclusion that the proposed method is reliable up to $x\approx 1$ with
a maximum error of about 5\%.

Probing the effect on the shape of the spectral densities by manually
varying single matrix elements  we find that the elements 
$(H_1+H_2)_{i,j}$ with $i,j\in \{1,2,3\}$ influence the line shapes
most. Naturally, matrix elements connecting short distances
$d$ have the biggest influence in a system with exponentially
decreasing correlation lengths. Thus, to further improve our results,
we replace these elements by the most reliable (Dlog-)Pad\'e
approximants of the underlying series expansions for each  $K$ considered.

\section{Summary}
In this article, the necessary details are given to understand how 
perturbative CUTs can be used to quantitatively calculate the
low-lying excitations of a certain class of many-particle
systems. Particular emphasis is put on spectral densities of
experimentally  
relevant observables. Due to the finiteness of the 
perturbation order, the method will work especially well for systems
with short-range correlations. Furthermore, it must be possible to
define suitable quasi-particles from which the whole spectrum of the
system under study can be constructed. The simplifications rendering
high order results for the eigen energies possible arise from mapping
the initial Hamiltonian onto an effective one which conserves the
number of particles. This enables separate calculations in the 0-, 1-,
2-,... particle sectors. In each sector, only a few-body problem has
to be solved. 

Effective observables representing
measurement processes are obtained in a similar fashion. In general,
they do not conserve the number of particles. But their action on
relevant states can be classified according to the number of particles
they inject into the system.

The antiferromagnetic spin 1/2 ladder is used to illustrate all steps of
the calculations in detail. The perturbation is taken about the limit
of isolated rungs. In this limit, rung-triplets are the elementary
excitations, which suggests to name them ``triplons''
(cf. Ref.~\cite{schmi03a}). They are suitable quasi-particles if the
inter-rung interaction is switched on inducing a magnetic 
polarization cloud.

The 0-, 1- and 2- particle sectors of the
effective Hamiltonian are studied separately. We address all possible
difficulties including a discussion of the finite clusters needed to
obtain the matrix elements for the infinite system by using
to the linked cluster theorem. 

Four different observables are discussed for the ladder system
relevant for neutron and light scattering experiments. We show in
detail how the relevant quantities can be calculated to obtain the
corresponding 1- and 2-triplon spectral densities for experiments at
zero temperature.  

The perturbative CUT methods requires the extrapolation of a large
number of quantities if the system is not very local. This is
especially so for calculations in the two- and more-particles sectors.
The article includes a general treatment of a novel
extrapolation scheme (optimized perturbation theory, OPT) designed to
simultaneously extrapolate a large number of quantities. The OPT is
introduced as a robust technique which does not necessarily render the best
possible results. But it provides reliable results for the 
regimes of interest. The method is
indispensable for situations where one needs automatized
extrapolations, a task that can hardly be solved by standard techniques
like Pad\'e methods.


\begin{thebibliography}{10}

\bibitem{gelfa00}
M.~P. Gelfand and R.~R.~P. Singh, Adv. Phys. {\bf 49},  93  (2000)

\bibitem{singh95}
R.~R.~P. Singh and M.~P. Gelfand, Phys. Rev. B {\bf 52},  R15695  (1995)

\bibitem{singh99}
R.~R.~P. Singh and Z. Weihong, Phys. Rev. B {\bf 59},  9911  (1999)

\bibitem{weiho01}
W. Zheng and J. Oitmaa, Phys. Rev. B {\bf 64},  014410  (2001)

\bibitem{uhrig98c}
G.~S. Uhrig and B. Normand, Phys. Rev. B {\bf 58},  R14705  (1998)

\bibitem{knett00b}
C. Knetter, A. B\"uhler, E. M\"uller-Hartmann, and G.~S. Uhrig, Phys. Rev.
  Lett. {\bf 85},  3958  (2000)

\bibitem{knett99a}
C. Knetter and G.~S. Uhrig, Eur. Phys. J. B {\bf 13},  209  (2000)

\bibitem{knett00a}
C. Knetter and G.~S. Uhrig, Phys. Rev. B {\bf 63},  94401  (2001)

\bibitem{knett00c}
C. Knetter, E. M\"uller-Hartmann, and G.~S. Uhrig, J. Phys.: Condens. Matter
  {\bf 12},  9069  (2000)

\bibitem{trebs00}
S. Trebst {\it et~al.}, Phys. Rev. Lett. {\bf 85},  4373  (2000)

\bibitem{weiho00a}
W. Zheng {\it et~al.}, Phys. Rev. B {\bf 63},  144410  (2001)

\bibitem{knett03a}
C. Knetter, K.~P.
 Schmidt, and G.~S. Uhrig, J. Phys. A: Math. Gen. {\bf 36},  7889   (2003)

\bibitem{knett01a}
C. Knetter, K.~P. Schmidt, M. Gr\"uninger, and G.~S. Uhrig, Phys. Rev. Lett.
  {\bf 87},  167204  (2001)

\bibitem{schmi01a}
K.~P. Schmidt, C. Knetter, and G. Uhrig, Europhys. Lett. {\bf 56},  877
  (2001)

\bibitem{knett02b}
C. Knetter, K.~P. Schmidt, and G.~S. Uhrig, Physica B {\bf 312},  527  (2002)

\bibitem{windt01}
M. Windt {\it et~al.}, Phys. Rev. Lett. {\bf 87},  127002  (2001)

\bibitem{gruni02c}
M. Gr\"uninger {\it et~al.}, Physica B {\bf 312},  617  (2002)

\bibitem{schmi03a}
K.~P. Schmidt and G. Uhrig, Phys. Rev. Lett. {\bf 90},  227204  (2002)

\bibitem{schmi03b}
K.~P. Schmidt, C. Knetter, M. Gr\"uninger, and G. Uhrig, Phys. Rev. Lett. {\bf
  90},  167201  (2003)

\bibitem{schmi03c}
K.~P. Schmidt, C. Knetter, and G. Uhrig, cond-mat/0307678,
Phys. Rev. B in press

\bibitem{knett03b}
C. Knetter and G.~S. Uhrig, Phys. Rev. Lett. {\bf 92}, 027204 (2004).

\bibitem{zheng03a}
W. Zheng, C.J. Hamer and R.R.P. Singh,
Phys. Rev. Lett. {\bf 91}, 037206   (2003)

\bibitem{barne93}
T. Barnes, E. Dagotto, J. Riera, and E.~S. Swanson, Phys. Rev. B {\bf 47},
  3196  (1993)

\bibitem{barne94}
T. Barnes and J. Riera, Phys. Rev. B {\bf 50},  6817  (1994)

\bibitem{gopal94}
S. Gopalan, T.~M. Rice, and M. Sigrist, Phys. Rev. B {\bf 49},  8901  (1994)

\bibitem{totsu95}
K. Totsuka and M. Suzuki, J. Phys.: Condens. Matter {\bf 7},  6079  (1995)

\bibitem{brehm96}
S. Brehmer, H. Mikeska, and U. Neugebauer, J. Phys.: Condens. Matter {\bf 8},
  7161  (1996)

\bibitem{weiho96a}
J.Oitmaa, R.~R.~P. Singh, and Z. Weihong, Phys. Rev. B {\bf 54},  1009  (1996)

\bibitem{sylju97}
O.~F. Syljuasen, S. Chakravarty, and M. Greven, Phys. Rev. Lett. {\bf 78},
  4115  (1992)

\bibitem{sush98}
O.~P. Sushkov and V.~N. Kotov, Phys. Rev. Lett. {\bf 81},  1941  (1998)

\bibitem{natsu98}
Y. Natsume, Y. Watabe, and T. Suzuki, J. Phys. Soc. Jpn. {\bf 67},  3314
  (1998)

\bibitem{weiho98a}
W.~H. Zheng, V. Kotov, and J. Oitmaa, Phys. Rev. B {\bf 57},  11439  (1998)

\bibitem{pieka98}
J. Piekarewicz and J.~R. Shepard, Phys. Rev. B {\bf 57},  10260  (1998)

\bibitem{brehm99}
S. Brehmer {\it et~al.}, Phys. Rev. B {\bf 60},  329  (1999)

\bibitem{kotov99}
V.~N. Kotov and O.~P. Sushkov, Phys. Rev. B {\bf 59},  6266  (1999)

\bibitem{breni00}
C. Jurecka and W. Brenig, Phys. Rev. B {\bf 61},  14307  (2000)

\bibitem{singh00}
P.~J. Freitas and R.~R.~P. Singh, Phys. Rev. B {\bf 62},  14113  (2000)

\bibitem{zvyag01}
A.~A. Zvyagin, J. Phys. A: Math. Gen. {\bf 34},  R21  (2001)

\bibitem{johns00}
D. Johnston {\it et~al.}, cond-mat/0001147.

\bibitem{kojim95}
K. Kojima {\it et~al.}, Phys. Rev. Lett. {\bf 74},  2812  (1995)

\bibitem{schwe96}
H. Schwenk {\it et~al.}, Solid State Commun. {\bf 100},  381  (1996)

\bibitem{eccle96}
R.~S. Eccleston, M. Azuma, and M. Takano, Phys. Rev. B {\bf 53},  14721
  (1996)

\bibitem{kuma97}
K. Kumagai, S. Tsuji, M. Kato, and Y. Koike, Phys. Rev. Lett. {\bf 78},  1992
  (1997)

\bibitem{hamma97}
P.~R. Hammar, D.~H. Reich, C. Broholm, and F. Trouw, Phys. Rev. B {\bf 57},
  7846  (1995)

\bibitem{sugai99}
S. Sugai and M. Suzuki, Phys. Stat. Sol. (b) {\bf 215},  653  (1999)

\bibitem{matsu00b}
M. Matsuda {\it et~al.}, prb {\bf 62},  8903  (2000)

\bibitem{konst01}
M.~J. Konstantinovic, J.~C. Irwin, M. Isobe, and Y. Ueda, Phys. Rev. B {\bf
  65},  12404  (2002)

\bibitem{uehar96}
M. Uehara {\it et~al.}, J. Phys. Soc. Jpn. {\bf 65},  2764  (1996)

\bibitem{steve81}
P.~M. Stevenson, Phys. Rev. D {\bf 23},  2916  (1981)

\bibitem{homepages}
  www.thp.uni-koeln.de/\~\ \!ck or
  www.thp.uni-koeln.de/\~\ \!gu 

\bibitem{heidb02b}
C. Heidbrink and G. Uhrig, Eur. Phys. J. B {\bf 30},  443  (2002)

\bibitem{dagot96}
E. Dagotto and T.~M. Rice, Science {\bf 271},  618  (1996)

\bibitem{shelt96}
D.~G. Shelton, A.~A. Nersesyan, and A.~M. Tsvelik, Phys. Rev. B {\bf 53},  8521
   (1996)

\bibitem{greve96}   
M. Greven, R.~J. Birgeneau, and U.-J. Wiese,
Phys. Rev. Lett. {\bf 77}, 1865 (1996)
 
\bibitem{domb83_pade}
{\em Phase Transitions and Critical Phenomena}, edited by C. Domb and J.~L.
  Lebowitz (Academic Press, London, 1983), Vol.~13

\bibitem{schmi03e}
K.~P. Schmidt, H. Monien, and G.~S. Uhrig, Phys. Rev. B {\bf 67},
184413 (2003)  

\bibitem{schmi02}
K.~P. Schmidt, C. Knetter, and G. Uhrig, Acta Physica Polonica B {\bf 34},
  1481  (2003)

\bibitem{zwanz61}
R. Zwanzig,  in {\em Lectures in Theoretical Physics}, edited by W.~E. Brittin,
  B.~W. Downs, and J.Downs (Interscience, New York, 1961), Vol.~III

\bibitem{mori65b}
H. Mori, Prog. Theor. Phys. {\bf 34},  399  (1965)

\bibitem{gagli87}
E.~R. Gagliano and C.~A. Balseiro, Phys. Rev. Lett. {\bf 26},  2999  (1987)

\bibitem{petti85}
D.~G. Pettifor and D.~L. Weaire, {\em The Recursion Method and its
  Applications}, Vol.~58 of {\em Springer Series in Solid State Sciences} (D.
  G. Pettifor and D. L. Weaire, Berlin, 1985)

\bibitem{viswa94}
V.~S. Viswanath and G. M\"uller, {\em The Recursion Method; Application to
  Many-Body Dynamics}, Vol.~m23 of {\em Lecture Notes in Physics}
  (Springer-Verlag, Berlin, 1994)

\bibitem{bende69}
  C. Bender and T. Wu, Phys. Rev. {\bf 184},  1231  (1969)
  

\end{thebibliography}

\end{document}